\documentclass[onecolumn,amsmath,showkeys,amssymb]{revtex4}
\usepackage{graphicx, subfigure}
\usepackage{amssymb, amsmath,amssymb,amsfonts}
\usepackage{amsthm,mathrsfs,amsopn}
\usepackage{dcolumn}
\usepackage{bm}
\usepackage{color}
\usepackage[utf8]{inputenc}
\usepackage{mathtools}

\theoremstyle{plain} %% This is the default

\newtheorem{theorem}{Theorem}

\newtheorem{remark}[theorem]{Remark}

\begin{document}

\title{Theory of synchronisation and pattern formation on time varying networks}

\author{Timoteo Carletti}
\email{timoteo.carletti@unamur.be}
\address{Department of Mathematics and naXys, Namur Institute for Complex Systems, University of Namur, Rue Graf\'e 2, B5000 Namur, Belgium}

\author{Duccio Fanelli}
\email{duccio.fanelli@unifi.it}
\address{Dipartimento di Fisica e Astronomia, Universit\`a di Firenze, INFN and CSDC,  Via Sansone 1, 50019 Sesto Fiorentino, Firenze, Italy}

\begin{abstract} 
Synchronisation and pattern formation have been intensely addressed for systems evolving on static networks. Extending the study to include the inherent ability of the network to adjust over time proved cumbersome and led to conclusions which lack of generality, as relying on peculiar  assumptions. Here, the master stability formalism is extended to account, in a thoroughly general prospect, for the additional contributions as stemming from the time evolution of the underlying network. The theory is successfully challenged against two illustrative testbeds, 
which can be respectively ascribed to synchronisation and Turing settings. 
\end{abstract}

%\pacs{89.75.Hc, 89.75.Kd,89.75.Fb}

\maketitle

%\vspace{0.8cm}

\section{Introduction}
\label{sec:intro}
Networks define versatile mathematical tools which can be conveniently invoked for modelling a plethora of physical systems~\cite{AlbertBarabasi,newmanbook}.  Local interactions among elementary constituents take place on isolated patches, the nodes of the network, while long-ranged exchanges crawls across the links, typically driven by diffusion, bridging thereby adjacent nodes of the collection.  Networks of coupled oscillators can self-organise to operate in unison~\cite{Pikovsky2001,arenasreview}, a condition that is necessary to sustain normal brain activity but which can yield pathological states in case of hyper-synchronisation, i.e., the inability of neurones to desynchronise~\cite{AECPlos2018}. In a completely different context, favouring a perfect synchrony of cyclic rhythms is vital for an optimal handling of energy production and distribution in power grids~\cite{MMAN2013}. Moreover, spatial non homogenous stationary stable motifs can spontaneously emerge following a symmetry breaking mechanism~\cite{Turing}, the analogue of a Turing instability for reaction-diffusion systems anchored on networks~\cite{NM2010}, which amplifies small perturbations acting on a uniform background. 

These phenomena have been mainly characterised with reference to static networks. That is, the network connections do not change over time, or, alternatively, the rate of modulation is very slow as compared to the typical time scales that regulate the dynamics of the state variables. The few attempts reported in the literature to generalise beyond this setting, focused either on fast switching networks~\cite{SBGR2006,PABFC2017,LFCP2018} among distinct static configurations, thus implying that the natural time scale of the network evolution is extremely fast as compared to that associated to the underlying dynamical system. In~\cite{AmritkarHu2006,BHCAKP,ZS2021} commutative temporal networks are instead invoked, a working hypothesis which may prove too restrictive because it assumes the eigenvectors of the relevant involved matrices, e.g., the Laplacian or the adjacency matrix, to not evolve in time. Other attempts to tackle the problem of a comprehensive theory of pattern formation on time dependent networks do not bear the sought generality, because they are once again implicitly based on the assumption of static, or very slowly varying, Laplace eigenvectors~\cite{vangorder2}, as we will show in the following.

 All endeavours must however deal with revisiting the seminal concept of master stability equation~\cite{Pecora}, so as to incorporate an explicit account of the inherent plasticity of the underlying network~\cite{Holme2013,MR2016}. The intent is accomplished in this work, where the master stability formalism is expanded to include the contributions stemming from an imposed time evolution of the links weight. The theory is tested against two distinct applications designed to tackle both synchronisation and Turing settings. The general framework that we shall here address assumes (nonlinear) diffusive inter-node exchanges. This is a mandatory pre-requisite for a uniform solution of the extended system to exist and, as such, it is routinely invoked  in different realms. Another viable scenario is found when the coupling term is a nonlinear function of the difference of the state variables referred to connected nodes (e.g., the paradigmatic Kuramoto model~\cite{Kuramoto} and its extension). This falls also under the umbrella of the developed theory, as we shall hereafter argue. As we will show the theory hereby developed confirms that synchronisation can be  enhanced by commutative temporal networks as reported in the litterature~\cite{BHCAKP,GFRMRABB2022}; we moreover take a step forward by showing that such a claim holds true beyond the limited setting of commuting networks. Similarly, we also show that the constraint on the curvature of the master stability function recently introduced in~\cite{ZS2021} can be relaxed once we remove the assumption of commutative temporal networks; indeed there exist time varying networks whose dynamics synchronise for a larger interval of the coupling strength without meeting the above condition. Finally, the proposed general framework contributes to significantly expand our current understanding on Turing patterns on time varying networks, beyond the settings so far considered which relied on the fast switching assumption~\cite{PABFC2017} or on peculiar choices of the evolution of the eigenvectors of the Laplace matrix~\cite{vangorder2}.

\section{Materials and Methods} 
\label{sec:themodel}
To set the stage for further analysis, we will begin our discussion by inspecting the spectral characteristics of a time dependent discrete Laplacian operator. Consider a symmetric time varying network made by $n$ nodes, characterised by the adjacency matrix $\mathbf{A}(t)$:  $A_{ij}(t)=A_{ji}(t)\neq 0$ whenever nodes $i$ and $j$ are connected at time $t$ and $A_{ij}(t)=A_{ji}(t)=0$ otherwise. Given the adjacency matrix one can construct the (combinatorial) Laplace matrix, $L_{ij}(t)= A_{ij}(t)-\delta_{ij} k_i(t)$, where $k_i(t)=\sum A_{ij}(t)$ denotes the degree of node $i$ at time $t$ and $\delta_{ij}$ is the Kronecker-$\delta$. Since $\mathbf{L}(t)$ is a symmetric matrix, for all $t$, one can find a (time dependent) orthonormal basis of eigenvectors, $\vec{\phi}^{(\alpha)}(t)$, associated with the eigenvalues $\Lambda^{(\alpha)}(t)\leq 0$ such that 
\begin{equation*}
\mathbf{L}(t)\vec{\phi}^{(\alpha)}(t)=\Lambda^{(\alpha)}(t)\vec{\phi}^{(\alpha)}(t) \quad \forall \alpha=1,\dots, n \text{ and } \forall t\, .
\end{equation*}
Moreover 
 \begin{equation}
 \left(\vec{\phi}^{(\alpha)}(t)\right)^\top\cdot \vec{\phi}^{(\beta)}(t)=\delta_{\alpha \beta}\, ,
\label{eq:eigeq2}
\end{equation}
where the dot represents  the scalar product and $\left( \right) ^\top$ stands for the transpose operation. Finally, with no loss of generality, we order the eigenvalues in such a way that $0=\Lambda^{(1)}(t)>\Lambda^{(j)}(t)$ for all $j\in\{2,\dots,n\}$ and we recall that $\vec{\phi}^{(1)}(t)=(1,\dots,1)^\top/\sqrt{n}$.

Assume that the eigenvectors evolve smoothly in time. Then, one can express the eigenvectors change rate as:
\begin{equation}
\label{eq:cab}
\frac{d \vec{\phi}^{(\alpha)}}{dt}(t)=\sum_{\beta}c_{\alpha\beta}(t)\vec{\phi}^{(\beta)}(t)\quad\forall \alpha=1,\dots, n\, .
\end{equation}
where $\mathbf{c}(t)$ is a $n\times n$ time dependent matrix that quantifies the projections on the independent eigendirections. By recalling the orthonormality condition~\eqref{eq:eigeq2} we can straightforwardly conclude that $\mathbf{c}$ is a real skew symmetric matrix with a null first row and first column, i.e., $c_{\alpha\beta}+c_{\beta\alpha}=0$ and $c_{1\alpha}=0$.

The time evolution of the eigenvectors is hence self-consistently ruled by the system of ODEs~\eqref{eq:cab}, complemented by the initial conditions, i.e., the Laplace eigenbasis at $t=0$. Notice that the case of switching networks can be also brought back to the above scenario, by approximating piecewise regular functions with smooth profiles or, alternatively, using the Fourier transform (see Appendix~\ref{sec:smallnet}).
Finally, we require the eigenvalues to satisfy standard conditions ensuring the existence and uniqueness of the linear system~\eqref{eq:GLHGlinalpha3}, so in particular differentiability is no longer required as in~\cite{vangorder2}.

We are now in a position to  elaborate on the conditions that generalise the master stability formalism to systems defined on time varying networks. To this end, consider a $d$-dimensional system described locally, i.e., at each node,  by the following ODE:
\begin{equation}
\label{eq:dotxF}
\frac{d\mathbf{x}}{dt}=\mathbf{F}(\mathbf{x})\quad \mathbf{x}\in\mathbb{R}^d\, ,
\end{equation}
where $\mathbf{F}$ is an arbitrary nonlinear function. Further, assume $n$ identical copies of the above system to be coupled via a time varying network through diffusive interactions modified with the inclusion of a nonlinear function $\mathbf{H}$:
\begin{equation}
\label{eq:maineq}
\frac{d\mathbf{x}_i}{dt}=\mathbf{F}(\mathbf{x}_i) +\varepsilon \sum_{j} L_{ij}(t) \mathbf{H}(\mathbf{x}_j)\, ,
\end{equation}  
where $\mathbf{x}_i=(x_i^{(1)},\dots,x_i^{(d)})^\top$ photographs the state of the system on node $i$, $\varepsilon>0$ is the strength of the coupling and $L_{ij}(t)$ are the entries of the time varying Laplace matrix.

Let us now fix a reference orbit, $\mathbf{s}(t)$, of the aspatial system~\eqref{eq:dotxF}. By exploiting the obvious condition $\sum_j L_{ij}(t)=0$ for all $i=1,\dots, n$ and all $t$, it is immediate to conclude that $\mathbf{s}(t)$ is also solution of Eq.~\eqref{eq:maineq}. Namely the coupled system exhibits a spatially homogeneous synchronous solution. Assuming the latter solution to be stable for the decoupled system~\eqref{eq:dotxF}, the question to be answered is whether it can turn unstable (or conversely, preserve its stability) when the inter-nodes couplings get activated by a small heterogeneous perturbation.  Denote by $\delta\mathbf{x}_i=\mathbf{x}_i-\mathbf{s}$ the deviations from the reference orbit and by assuming these latter small, one can derive a self-consistent set of linear ODE for tracking the evolution of the perturbation in time. To this end,  we introduce $\delta\mathbf{x}_i$ in Eq.~\eqref{eq:maineq} and perform a Taylor expansion arrested to linear order, to eventually get:
\begin{equation}
\label{eq:GLHGlin}
\frac{d\delta\mathbf{x}_i}{dt}=\mathbf{J}_\mathbf{F}(\mathbf{s}(t))\delta\mathbf{x}_i +\varepsilon \sum_{j} {L}_{ij}(t) \mathbf{J}_\mathbf{H}(\mathbf{s}(t))\delta\mathbf{x}_j\, , 
\end{equation}
where $\mathbf{J}_\mathbf{F}(\mathbf{s}(t))$ (resp. $\mathbf{J}_\mathbf{H}(\mathbf{s}(t))$) denotes the Jacobian matrix of the function $\mathbf{F}$ (resp. $\mathbf{H}$) evaluated on the trajectory $\mathbf{s}(t)$. Remark that a completely equivalent governing equation is obtained when starting from an inter-nodes coupling of the type $\sum_j A_{ij} \mathbf{H}(\mathbf{x}_j-\mathbf{x}_i)$, with $\mathbf{H}(\mathbf{0})=\mathbf{0}$, as anticipated above.

To make further progress in the study of the linear non-autonomous system (\ref{eq:GLHGlin}), we project $\delta\mathbf{x}_i$ onto the orthonormal basis formed by the eigenvectors of $\mathbf{L}(t)$, to yield $\delta\mathbf{x}_i=\sum_\alpha \delta\hat{\mathbf{x}}_{\alpha}\phi^{(\alpha)}_i$. By inserting the latter into~\eqref{eq:GLHGlin} and recalling the definition of matrix $\mathbf{c}(t)$, one obtains for all $\beta$ (see Appendix~\ref{sec:MSEtv}):
\begin{equation}
\label{eq:GLHGlinalpha3}
\frac{d\delta\hat{\mathbf{x}}_{\beta}}{dt} = \sum_\alpha c_{\beta\alpha}(t)\delta\hat{\mathbf{x}}_{\alpha}+\left[\mathbf{J}_\mathbf{F}(\mathbf{s}(t))+\varepsilon \Lambda^{(\beta)}(t)\mathbf{J}_\mathbf{H}(\mathbf{s}(t))\right]\delta\hat{\mathbf{x}}_{\beta}\, . 
\end{equation}

 By introducing  $\delta\hat{\mathbf{x}}=(\delta\hat{\mathbf{x}}^\top_{1},\dots,\delta\hat{\mathbf{x}}_{n}^\top)^\top$ one can cast Eq.~\eqref{eq:GLHGlinalpha3} in  compact form:
  \begin{equation}
\label{eq:GLHGlinalpha3compact}
\frac{d\delta\hat{\mathbf{x}}}{dt} =  \left[\mathbf{c}\otimes \mathbf{1}_d+\mathbf{1}_n\otimes \mathbf{J}_\mathbf{F}+\varepsilon \mathbf{\Lambda}\otimes\mathbf{J}_\mathbf{H}\right]\delta\hat{\mathbf{x}}:=\mathbf{M}\delta\hat{\mathbf{x}}\, ,
\end{equation}
where $\otimes$ denotes the Kronecker product, $\mathbf{\Lambda}=\mathrm{diag}\left(\Lambda^{(1)},\dots,\Lambda^{(n)}\right)$ and, given any positive integer $m$, $\mathbf{1}_m$ denotes the $m$ dimensional identity matrix. Let us observe that the latter formula and the following analysis differ from the one presented in~\cite{vangorder2} where the perturbation is assumed to align onto a single mode, hypothesis that ultimately translates in the stationarity of the Laplace eigenvectors, that is $\mathbf{c}=\mathbf{0}$. The same assumption is also at the root of the results by~\cite{ZS2021}; indeed commuting time varying networks implies to deal with a constant eigenbasis. In conclusion, Eq.~\eqref{eq:GLHGlinalpha3} or Eq.~\eqref{eq:GLHGlinalpha3compact} are capable to describe the projection of the linearised dynamics on a generic time varying Laplace eigenbasis, and thus allow us to draw general conclusions without simplifying assumptions.

The above system~\eqref{eq:GLHGlinalpha3} represents the generalised version of the celebrated Master Stability Equation (MSE)~\cite{Pecora,HCLP}, which includes an explicit account of the network evolution as encoded in matrix $\mathbf{c}(t)$, as well as in the dependence of $\Lambda^{(\beta)}(t)$ against time. Its largest Lyapunov exponent quantifies the exponential rate at which an infinitesimal perturbation in the transverse subspace grows: it defines an improved version of the Master Stability Function (MSF) and enables one to draw conclusions about the stability of the reference orbit. In concrete terms, consider the matrix equation $d \mathbf{O} / dt = \mathbf{M} \mathbf{O}$ where $\mathbf{O}(0)=\mathbf{1}_{nd}$. Solve the preceding equation numerically and compute $\nu_i(t)$ ($i=1,..., nd$)  the (time dependent) eigenvalues of $\mathbf{O}(t)$. The Lyapunov 
exponents are the computed by $\lambda_i=\lim_{t \rightarrow \infty} \ln \nu_i / t$.  Notice that~\eqref{eq:GLHGlinalpha3compact} -- or its equivalent counterpart~\eqref{eq:GLHGlinalpha3} from which matrix $\mathbf{M}$ originates -- displays two independent time scales (in addition to the ones characterising the reactive dynamics), one reflecting  the dynamics of the isolated units, i.e., $\mathbf{s}(t)$, and the other stemming from the network modulation over time, as mirrored in $\mathbf{c}(t)$ and $\Lambda^{(\beta)}(t)$. Hence, the examined system cannot be managed via standard Floquet methods, not even when periodic homogeneous $\mathbf{s}(t)$ are concerned. Further, classical MSF approaches -- carried out over static networks -- can be conveniently simplified by considering the evolution of the imposed perturbation along each independent direction (the eigenvectors), associated to different eigenvalues of the Laplacian. This path cannot be pursued here as the matrix $\mathbf{c}(t)$ is responsible for a non trivial entanglement of different modes, as also remarked in~\cite{vangorder3}. In general, system (\ref{eq:GLHGlinalpha3compact}) should be hence handled numerically (see e.g.,~\cite{HCLP} for an account of the subtleties to be faced when carrying the numerical computation). 

For demonstrative purposes and to highlight the consequences resulting from the form of Eq.~\eqref{eq:GLHGlinalpha3}, we will hereby concentrate onto two illustrative examples, which are constructed so as to yield a time independent generalised MSF. In both cases, the network is made of just three nodes, the eigenvalues are assumed constant and the time derivative of the eigenvectors  projected on the eigenbasis returns a matrix $\mathbf{c}$ that does not change in time (details are supplied in~\ref{sec:smallnet}). In the first example, the reference orbit is stationary and we will operate in the furrow of the Turing  instability~\cite{Turing,NM2010}. In the second case, we will assume a set of nonlinearly coupled Stuart-Landau (SL) oscillators~\cite{vanharten,aranson,garcamorales}, to investigate the impact of the network dynamics on the onset of synchronisation.

\section{Turing instability on time varying network}
\label{sec:turing} 
The reference solution $\mathbf{s} \equiv \mathbf{s}_0$ is assumed stationary and the coupling linear, i.e., $\mathbf{H}(\mathbf{x})=\mathbf{D}\mathbf{x}$, where $\mathbf{D}$ is a suitable diagonal matrix with positive entries. To reduce to the usual Turing setting we posit $d=2$, namely $\mathbf{x}_i=(u_i,v_i)$, $\mathbf{F}(\mathbf{x}_i)=(f(u_i,v_i),g(u_i,v_i))$ and $\mathbf{D}=\mathrm{diag}(D_u,D_v)$. Then, Eq.~\eqref{eq:maineq} rewrites for all $i=1,\dots,n$
\begin{equation}
\label{eq:Tnet}
\begin{dcases}
\frac{du_i}{dt}&=f(u_i,v_i)+D_u\sum_{j=1}^{n}L_{ij}(t) u_j  \\ 
\frac{dv_i}{dt}&=g(u_i,v_i)+D_v\sum_{j=1}^{n}L_{ij}(t) v_j 
\end{dcases}\, ,
\end{equation}
where $D_u>0$ (resp. $D_v>0$) is the diffusion coefficients of species $u$ (reps. $v$) and ${L}_{ij}(t)$ are the elements of the above defined Laplace matrix of the time varying network. Recall that $\mathbf{s}_0=(u^*,v^*)$ is a stable equilibrium solution for the reaction part. Therefore, $f(u^*,v^*)=g(u^*,v^*)=0$, $\mathrm{tr}(\mathbf{J}_0)<0$ and $\det(\mathbf{J}_0)>0$, where $\mathbf{J}_0$ is the Jacobian of the reaction part evaluated at the equilibrium. Notice that $\mathbf{s}_0$ is also an equilibrium solution for the whole system of coupled equations~\eqref{eq:Tnet}. We are thus interested in studying its stability under an imposed node-dependent, hence heterogeneous  perturbation.

By setting $\delta x_i=u_i-u^*$, $\delta y_i=v_i-v^*$, one can follow the main steps of the theory presented above to eventually obtain the analogous of Eq.~\eqref{eq:GLHGlinalpha3compact},
\begin{equation}
\label{eq:Tnetlin5}
\frac{d \delta\hat{\mathbf{x}}}{dt}= \left[\mathbf{c}\otimes \mathbf{1}_2+\mathbf{1}_n\otimes \mathbf{J}_0+\mathbf{\Lambda}\otimes \mathbf{D}\right]\delta\hat{\mathbf{x}}\, ,
\end{equation}
where $\delta\hat{\mathbf{x}}=(\delta\hat{\mathbf{x}}_1^\top,\dots,\delta\hat{\mathbf{x}}_n^\top)^\top$ and  $\delta\hat{\mathbf{x}}_\beta=(\delta \hat{x}_{\beta},\delta \hat{y}_{\beta})$ represents the projection of the $i$-th perturbation $(\delta x_i,\delta y_i)$ on the $\beta$-eigenvector. 

For a definite application, we focus on the three-nodes time dependent network mentioned above and further characterised in~\ref{sec:smallnet}. For sake of definitiveness let us hereby recall that the matrix $\mathbf{c}$ defined by Eq.~\eqref{eq:cab} is given by
\begin{equation*}
 \mathbf{c}=\left(
\begin{matrix}
 0 & 0 & 0\\
 0 & 0 & \Omega\\
 0 & -\Omega& 0
\end{matrix}\right)\, ,
\end{equation*}
for some $\Omega >0$. That together with the following initial Laplace eigenbasis
\begin{equation*}
 \vec{\phi}^{(1)}(0)=\frac{1}{\sqrt{3}}\left(
\begin{smallmatrix}
 1\\1\\1
\end{smallmatrix}
\right)\, , \vec{\phi}^{(2)}(0)=\frac{1}{\sqrt{6}}\left(
\begin{smallmatrix}
 1\\-2\\1
\end{smallmatrix}
\right)\, ,\vec{\phi}^{(3)}(0)=\frac{1}{\sqrt{2}}\left(
\begin{smallmatrix}
 -1\\0\\1
\end{smallmatrix}
\right)
\end{equation*}
returns the following adjacency matrix
\begin{equation}
 A_{ij}(t)=\left(\begin{smallmatrix} 0 & \frac{1}{2}-\frac{\cos\left(\frac{\pi }{3}+2\Omega t\right)}{3} & \frac{\cos\left(2\Omega t\right)}{3}+\frac{1}{2}\\ \frac{1}{2}-\frac{\cos\left(\frac{\pi }{3}+2\Omega t\right)}{3} & 0 & \frac{1}{2}-\frac{\cos\left(\frac{\pi }{3}-2\Omega t\right)}{3}\\ \frac{\cos\left(2\Omega t\right)}{3}+\frac{1}{2} & \frac{1}{2}-\frac{\cos\left(\frac{\pi }{3}-2\Omega t\right)}{3} & 0 \end{smallmatrix}\right)\, .
\label{eq:adjt}
\end{equation}

Moreover, we assume the paradigmatic Brusselator scheme as the reference reaction model, namely $f(u,v)=1-(b+1)u+cu^2v$ and $g(u,v)=bu-cu^2v$, where $b$ and $c$ are the positive parameters. The stationary equilibrium is thus $u^* = 1$ and $v^* = b/c$, while the Jacobian of the reaction part evaluated on the equilibrium is $\partial_u f = b-1$, $\partial_vf = c$, $\partial_u g = -b$ and $\partial_v g = -c$. To exemplify our conclusion we will further set $D_u =0.01$ and $D_v = 1$,  compute the eigenvalues of the linear system~\eqref{eq:Tnetlin5} (here straightforward as the linear system has constant coefficients) and then characterise the region in the $(b,c)$ plane  where the instability is predicted to take place for (i) the standard Turing setting, i.e., when the network is made time independent ($\Omega=0$, which implies $\mathbf{c}$ to be the null matrix); (ii) the extended scenario when the network is made to evolve in time ($\Omega >0$, hence $\mathbf{c} \ne \mathbf{0}$).

\begin{figure*}[ht]
\centering
\includegraphics[scale=0.22]{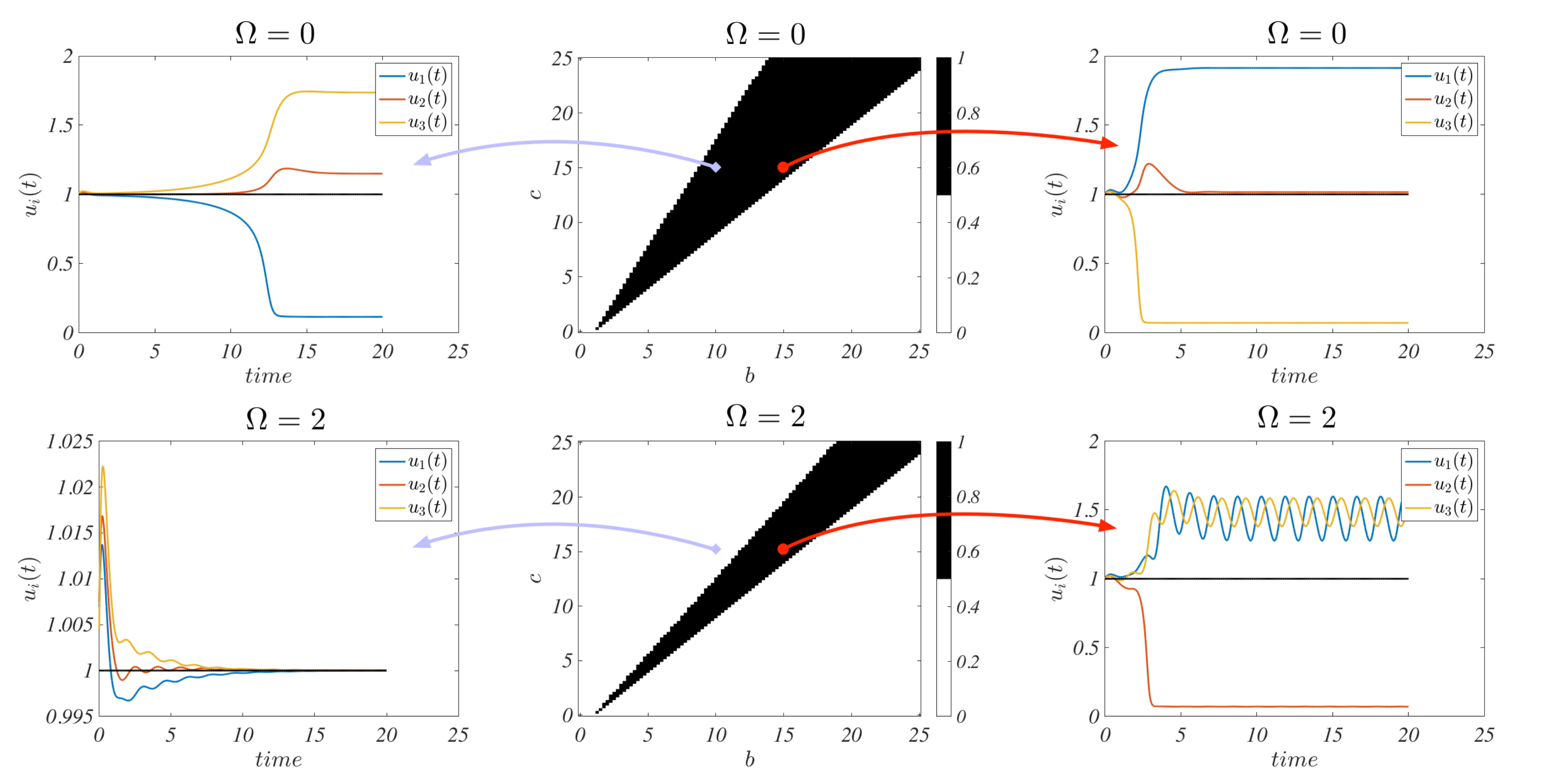}
%\vspace{-1.5cm}
\caption{\textbf{Turing instability on time varying networks.} Middle panels report the parameters region associated to the emergence of Turing instability (black region) for the Brusselator model, on respectively a static  (top panel) and time varying network (bottom panel, $\Omega=2$). The remaining panels display the computed trajectories for  $b=10$, $c=15$ (left, yielding to Turing patterns just for the case of a static network) and $b=c=15$ (right, patterns develop in both considered situations). Here, $D_u=0.01$ and $D_v=1.0$.}
\label{fig:resultsTuring}
\end{figure*}

The results reported in Fig.~\ref{fig:resultsTuring} show that, for this specific choice of the parameters, the region deputed to the instability (black shadow) shrinks when the system is made to evolve on a time varying network~\footnote{Throughout this work the numerical simulations have been performed using a $4$-th order Runge-Kutta scheme implemented in Matlab~\cite{MATLAB2021}. The initial conditions have been realised by drawing uniformly random perturbations $\delta$-close to the homogeneous equilibrium and the simulation time has been taken of the order of $(-\log \delta) / \rho_{\mathrm{MSF}}$, where $\rho_{\mathrm{MSF}}$ is the maximum of the MSF. This is indeed the time necessary to (possibly) increase the $\delta$-perturbation up to a macroscopic size. In the rest of the work we set $\delta=10^{-2}$, small enough to discriminate between the onset of the instability using a reasonable simulation time for the values of $\rho_{\mathrm{MSF}}$ we are dealing with.}. The outcome of the simulations corroborates the predictions, thus confirming the adequacy of the theory and the crucial role played by the extra contribution to the MSF which can be traced back to matrix $\mathbf{c}$.

\begin{figure*}[ht]
\centering
\includegraphics[scale=0.33]{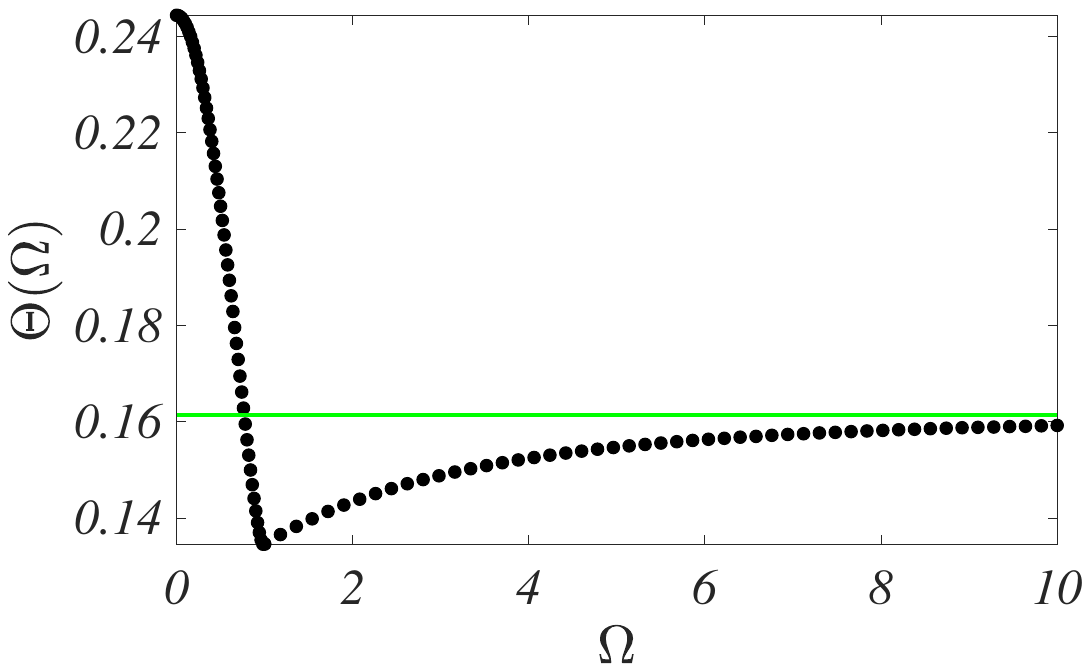}
\vspace{-3cm}
\caption{\textbf{Size of the Turing region with respect to $\Omega$.} For a fixed range of model parameters $(b,c)\in [0,25]\times[0,25]$ we report the size of the region associated to the emergence of Turing patterns for a given $\Omega$, assuming the network defined above as the underlying support for the dynamics. The horizontal green line stands for the size of the instability region as obtained for the averaged network. The diffusion coefficients have been set to $D_u=0.01$ and $D_v=1.0$.}
\label{fig:sizeTuring}
\end{figure*}

In Fig.~\ref{fig:sizeTuring} we report the size of the Turing instability region $\Theta(\Omega)$ as a function of $\Omega$ (black dots); by visual inspection one can appreciate that indeed the instability region shrinks once $\Omega$ increases, i.e., $\Theta(0)>\Theta(\Omega)$ for all $\Omega$. We can also observe  a non-monotone behaviour with a minimum for $\Omega=1$. The Figure also supports the correctness of the results proved in~\cite{PABFC2017}. Indeed,  for sufficiently fast network dynamics, the emergence of Turing pattern can be proved by looking at the behaviour of the system defined on the averaged network, whose associated size (i.e., the extension of the deputed region in the parameters plane) is given by the green line.

\section{Synchronisation of Stuart-Landau oscillators nonlinearly coupled via time varying networks}
\label{sec:SL}  
\begin{figure*}[ht]
\centering
\includegraphics[scale=0.22]{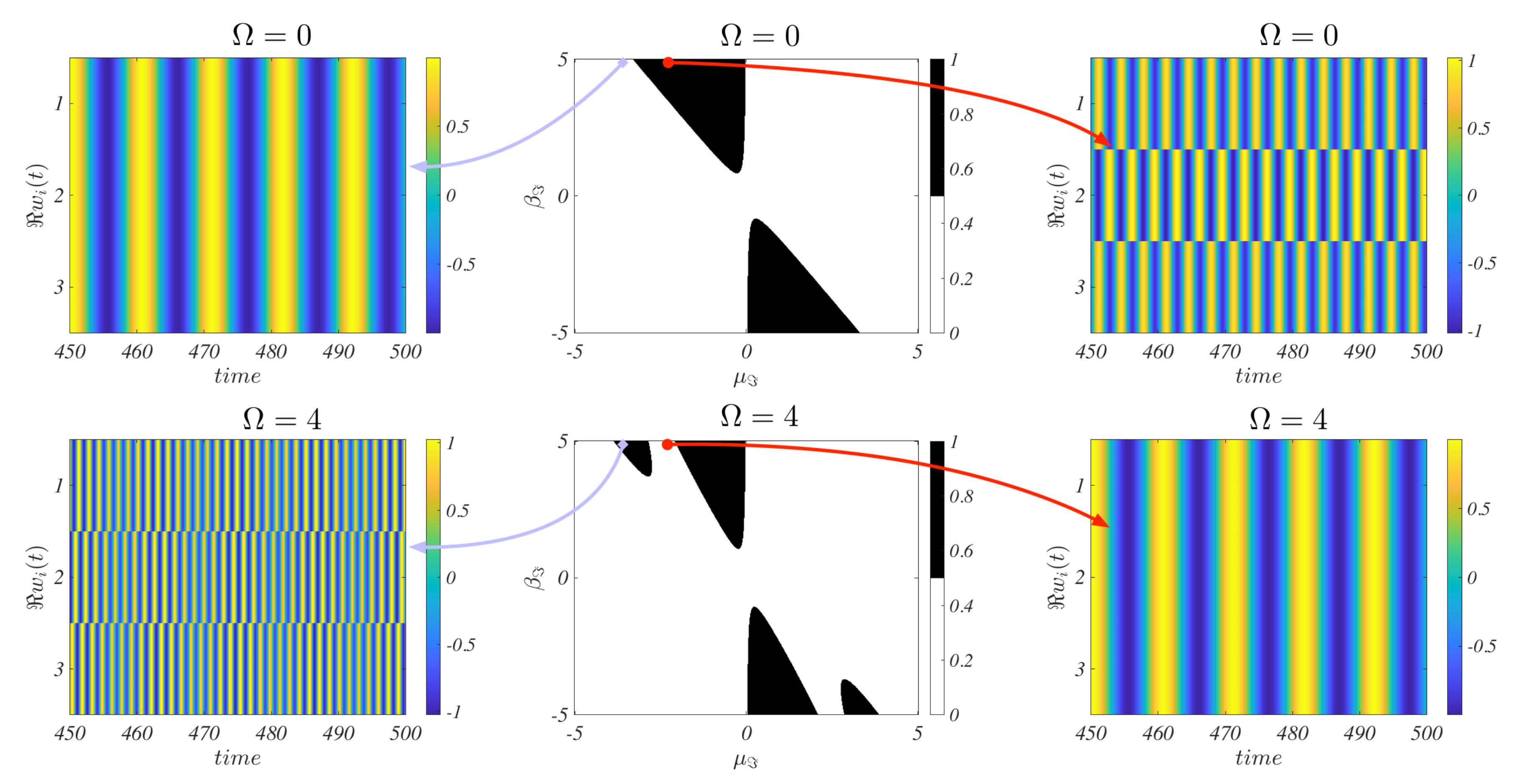}
%\vspace{-1.5cm}
\caption{\textbf{Synchronisation on time varying networks.} Middle panels identify the regions in the parameters plane $(\beta_\Im, \mu_\Im)$ where the synchronous solution is unstable (back domains) on, respectively, a static (top panel) and time varying network (bottom panel, $\Omega=4$). The remaining panels show the evolution in time of the real part of 
$w_j$, for different choices of the parameters (as indicated by the arrows). The two selected working points are $\beta_\Im=4.9$, $\mu_\Im=-2.3$ and $\beta_\Im=4.9$, $\mu_\Im=-3.6$. Here,  $\sigma = 1.0+4.3i$, $\beta_\Re = 1.0$, $\mu_\Re = 0.1$ and $m = 3$.}
\label{fig:resultsSynchro}
\end{figure*}

Consider a Stuart-Landau (SL) oscillator, the normal form for a generic system close to a supercritical Hopf-bifurcation. It is characterised by a complex amplitude $w$ which evolves in time according to $\dot{w}=\sigma w-\beta |w|^2w$, where $\sigma=\sigma_\Re+i\sigma_\Im$ and $\beta=\beta_\Re+i\beta_\Im$ are complex control parameters. The oscillator admits a limit cycle $\hat{z}(t)=\sqrt{\sigma_\Re/\beta_\Re}e^{i\omega t}$, where $\omega=\sigma_\Im-\beta_\Im \sigma_\Re/\beta_\Re$. The latter is a stable solution of an isolated SL equation, provided $\sigma_\Re>0$ and $\beta_\Re>0$, conditions that we hereby assume. To  proceed in the analysis we couple together 
$n$ identical SL oscillators, each bearing a complex amplitude $w_j$, with $j=1,...,n$:
\begin{equation}
\label{eq:maineqSL}
\frac{dw_j}{dt}= \sigma w_j-\beta w_j|w_j|^2+\mu \sum_{\ell} {L}_{j\ell}(t) H(w_\ell)\,  ,
\end{equation}
where $\mu=\mu_\Re+i\mu_\Im$ is a complex parameter that sets the strength of the coupling and where  $H(w)=w|w|^{m-1}$, for some integer $m\geq 1$. The above system falls in the class of Eq.~\eqref{eq:maineq}. We focus now on the extended solution $w_j=\hat{z}(t)$ $\forall j$ and inspect its stability, as a function of the model parameters, namely to the emergence of global synchronisation. To achieve this goal we set: 
\begin{equation}
\label{eq:wpert}
w_j(t)=\hat{z}(t)(1+\rho_j(t))e^{i\theta_j(t)}\, ,
\end{equation}
where the real functions $\rho_j(t)$ and $\theta_j(t)$ are assumed to be small. 
A straightforward computation (more details can be found in~\ref{sec:synchSL}) leads to :
\begin{equation}
\label{eq:maineqSLlinMSF}
\frac{d \delta\hat{\mathbf{x}}}{dt}= \left[\mathbf{c}\otimes \mathbf{1}_2+\mathbf{1}_n\otimes \mathbf{J}_0+\mathbf{\Lambda}\otimes \mathbf{J}_{H}\right]\delta\hat{\mathbf{x}}\, ,
\end{equation}
where $\delta\hat{\mathbf{x}}=(\delta\hat{\mathbf{x}}_1^\top,\dots,\delta\hat{\mathbf{x}}_n^\top)^\top$ and  where $\delta\hat{\mathbf{x}}_\beta=(\hat{\rho}_{\beta},\hat{\theta}_{\beta})$ denotes
the projection of $(\rho_i,\theta_i)$ on the $\beta$-eigenvector. The Jacobian of the isolated SL system is given by $\mathbf{J}_0=\left(
\begin{smallmatrix}
 -2\sigma_\Re & 0\\
  -2\beta_\Im \sigma_\Re/\beta_\Re & 0
\end{smallmatrix}\right)$, whereas the contributions stemming from the nonlinear coupling correspond to  $\mathbf{J}_H=\left(\frac{\sigma_\Re}{\beta_\Re}\right)^{\frac{m-1}{2}}\left(
\begin{smallmatrix}
m\mu_\Re & -\mu_\Im\\
m\mu_\Im & \mu_\Re
\end{smallmatrix}\right)$.
To illustrate the outcome of the analysis, we set $n=3$ and deal with the time dependent network introduced above (see also~\ref{sec:smallnet}). We let $\beta_\Im$ and $\mu_\Im$ to vary freely, and freeze the other parameters to nominal values, for the sake convenience. In Fig.~\ref{fig:resultsSynchro} we depict in black the region of the parameters plane $(\beta_\Im, \mu_\Im)$ where the generalised MSF is positive, i.e., where the uniform limit cycle solution is predicted to be unstable and thus synchronisation is excluded. The complementary domain points hence to the choice of the parameters that yields stable synchronous oscillations. The inherent dynamics of the network enhances the ability of the system to synchronise, at a global scale. Indeed the region in black shrinks when the network is made to evolve in time. However, the parameters  $\beta_\Im$ and $\mu_\Im$ can be chosen in such a way that the coupled collection of SL oscillators turns unstable when evolved on a dynamical support ($\Omega \ne 0$), while being stable on its static counterpart ($\Omega = 0$).

\begin{figure*}[ht]
\centering
\includegraphics[scale=0.33]{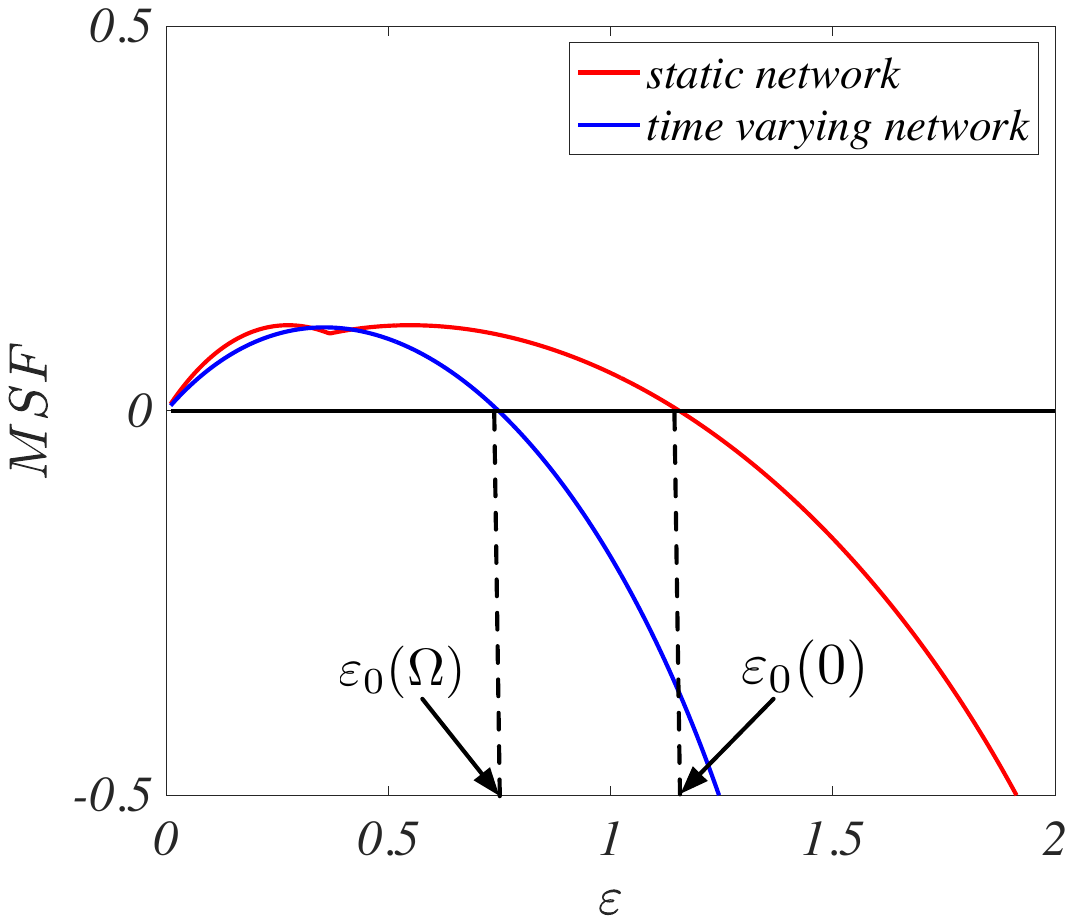}
\vspace{-2cm}
\caption{\textbf{Master Stability Function for the Stuart-Landau nonlinearly coupled oscillators.} We report the MSF computed for the SL system coupled with the simple $3$ nodes network as a function of $\varepsilon$. The remaining parameters have been fixed to the values, $\Omega=2.0$, $\sigma = 1.0+4.3i$, $\beta= 1.0+1.1i$, $\mu_0 = 0.1-0.5i$ and $m = 3$. We emphasised the first nontrivial zeros of the MSF,  $\varepsilon_0(\Omega)$, as as a function of the coupling strength and its dependence of $\Omega$.}
\label{fig:fig1SL}
\end{figure*}

To study the interplay of the coupling strength and the parameter $\Omega$ on synchronisation, let us set in Eq.~\eqref{eq:maineqSL} $\mu = \varepsilon \mu_0$ for a fixed complex parameter $\mu_0$ (hereby fixed, without loss of generality, to $\mu_0=0.1-0.5 i$) and a positive real parameter $\varepsilon$. In Fig.~\ref{fig:fig1SL} we report the behaviour of the MSF as a function of the coupling strength $\varepsilon$ for two values of $\Omega$ (blue curve $\Omega=2.0$, red curve $\Omega=0.0$, i.e., static network). Let us recall that values of $\varepsilon$ associated to a positive MSF correspond to an unstable limit cycle solution for the SL and thus to desynchronisation. The size of such interval of desynchronisation is given by the first nonzero value of $\varepsilon$ for which the MSF is zero, say $\varepsilon_0$. The latter clearly  depends on $\Omega$ and moreover $\varepsilon_0(2.0)>\varepsilon_0(0.0)$, namely the time varying network exhibits a larger domain of synchronisation, being the interval of instability smaller than in the static network case.

\begin{figure*}[ht]
\centering
\includegraphics[scale=0.33]{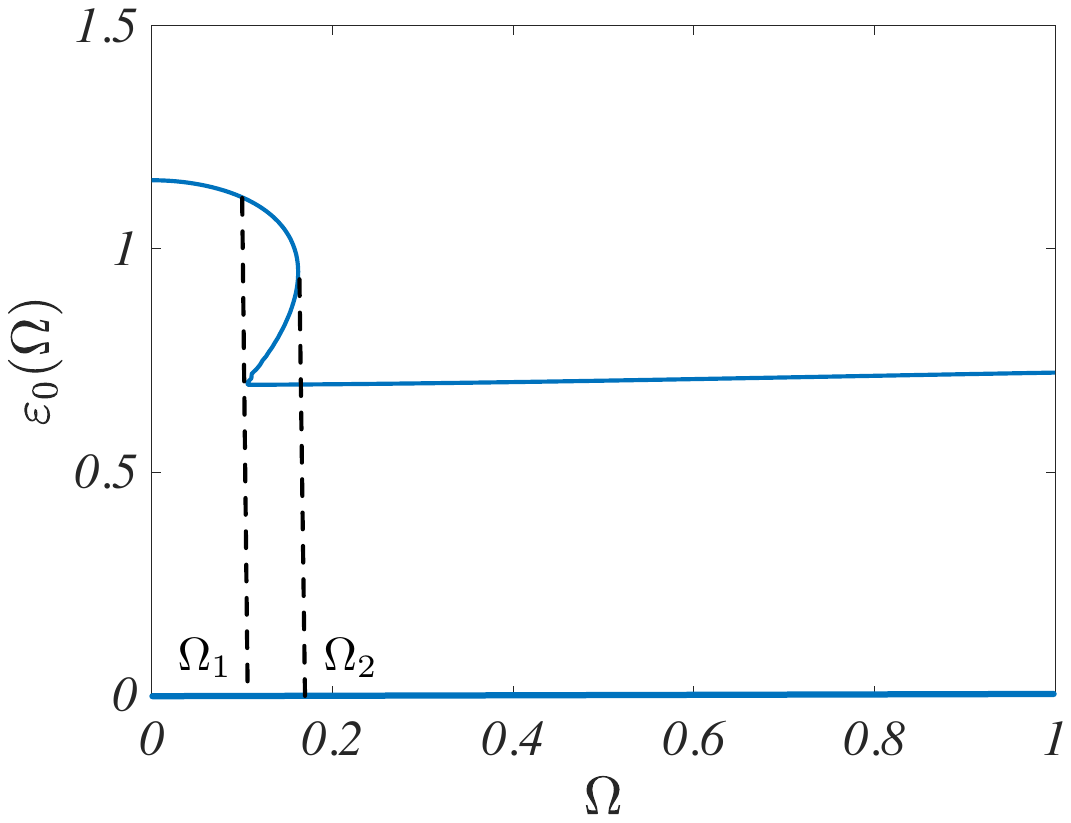}
\vspace{-2cm}
\caption{\textbf{Size of the synchronisation region for the Stuart-Landau nonlinearly coupled oscillators.} We report the value of $\varepsilon_0(\Omega)$ as a function of $\Omega$. The remaining parameters have been fixed to the values, $\sigma = 1.0+4.3i$, $\beta= 1.0+1.1i$, $\mu_0 = 0.1-0.5i$ and $m = 3$. Let us observe the existence of an interval of values $I=[\Omega_1,\Omega_2]$ such that for all $\Omega\in I$ there exist three values of $\varepsilon_0(\Omega)$ (see Appendix~\ref{sec:synchSL}).}
\label{fig:fig2SL}
\end{figure*}

The latter claim is further supported by the results displayed in Fig.~\ref{fig:fig2SL} where we report the dependence of $\varepsilon_0(\Omega)$ as a function of $\Omega$. One can observe that $\varepsilon_0(0)>\varepsilon_0(\Omega)$ for all the considered values of $\Omega$ and thus conclude that the time varying network exhibits a larger domain of synchronisation. We observe that such result is not limited to the small network here considered as numerically shown in~\ref{sec:smallnet}.

Let us remark that the MSF has a negative second derivative evaluated at $\varepsilon_0(\Omega)$ for all $\Omega$ (see Fig.~\ref{fig:fig1SL}), we can thus conclude that the result here presented generalises the one recently found in~\cite{ZS2021} without resorting to the restrictive assumption of commuting time varying networks: if the MSF is convex, with respect to the coupling strength, then the time varying network can synchronise for a range of $\varepsilon$ larger than that associated to its static analogue.

\section{Conclusions}
\label{sec:concl}
Summing up we have here extended the Master Stability theory to a setting where the time evolution of the underlying network support is explicitly accounted for. The dynamics of the network is reflected in a time dependent skew symmetric matrix $\mathbf{c}(t)$, that stems from the time evolution of the Laplacian eigenvectors. The corresponding eigenvalues can also adjust in time and contribute to shape the evolution of the imposed perturbation at a linear order of approximation. The proposed theory is general and applies to all systems for which the  Master Stability formalism was originally conceived. For illustrative purposes we have here decided to test it against two simple cases study, which can be respectively ascribed to synchronisation and Turing instability. The method applies however to more complex settings, as e.g., synchronisation of  chaotic trajectories (see Appendix~\ref{sec:synchLorenz}). In particular, we showed that the condition of the negative curvature of the MSF invoked in~\cite{ZS2021} to ensure that commuting time varying networks do synchronise, is no longer required when general time varying networks are accounted for. Indeed we have demonstrated that coupled Lorenz systems can easily synchronise once coupled with a time varying network, even if the associated MSF has a positive curvature (see Appendix~\ref{sec:synchLorenz}). 

\vspace{0.5cm}
\noindent\textbf{Acknowledgement.} We thank Robert A Van Gorder for fruitful discussions.

 \bibliographystyle{apsrev4-1}
\bibliography{bib_RTP}

%merlin.mbs apsrev4-1.bst 2010-07-25 4.21a (PWD, AO, DPC) hacked
%Control: key (0)
%Control: author (72) initials jnrlst
%Control: editor formatted (1) identically to author
%Control: production of article title (-1) disabled
%Control: page (0) single
%Control: year (1) truncated
%Control: production of eprint (0) enabled
\begin{thebibliography}{29}%
\makeatletter
\providecommand \@ifxundefined [1]{%
 \@ifx{#1\undefined}
}%
\providecommand \@ifnum [1]{%
 \ifnum #1\expandafter \@firstoftwo
 \else \expandafter \@secondoftwo
 \fi
}%
\providecommand \@ifx [1]{%
 \ifx #1\expandafter \@firstoftwo
 \else \expandafter \@secondoftwo
 \fi
}%
\providecommand \natexlab [1]{#1}%
\providecommand \enquote  [1]{``#1''}%
\providecommand \bibnamefont  [1]{#1}%
\providecommand \bibfnamefont [1]{#1}%
\providecommand \citenamefont [1]{#1}%
\providecommand \href@noop [0]{\@secondoftwo}%
\providecommand \href [0]{\begingroup \@sanitize@url \@href}%
\providecommand \@href[1]{\@@startlink{#1}\@@href}%
\providecommand \@@href[1]{\endgroup#1\@@endlink}%
\providecommand \@sanitize@url [0]{\catcode `\\12\catcode `\$12\catcode
  `\&12\catcode `\#12\catcode `\^12\catcode `\_12\catcode `\%12\relax}%
\providecommand \@@startlink[1]{}%
\providecommand \@@endlink[0]{}%
\providecommand \url  [0]{\begingroup\@sanitize@url \@url }%
\providecommand \@url [1]{\endgroup\@href {#1}{\urlprefix }}%
\providecommand \urlprefix  [0]{URL }%
\providecommand \Eprint [0]{\href }%
\providecommand \doibase [0]{http://dx.doi.org/}%
\providecommand \selectlanguage [0]{\@gobble}%
\providecommand \bibinfo  [0]{\@secondoftwo}%
\providecommand \bibfield  [0]{\@secondoftwo}%
\providecommand \translation [1]{[#1]}%
\providecommand \BibitemOpen [0]{}%
\providecommand \bibitemStop [0]{}%
\providecommand \bibitemNoStop [0]{.\EOS\space}%
\providecommand \EOS [0]{\spacefactor3000\relax}%
\providecommand \BibitemShut  [1]{\csname bibitem#1\endcsname}%
\let\auto@bib@innerbib\@empty
%</preamble>
\bibitem [{\citenamefont {Albert}\ and\ \citenamefont
  {Barab{\'a}si}(2002)}]{AlbertBarabasi}%
  \BibitemOpen
  \bibfield  {author} {\bibinfo {author} {\bibfnamefont {R.}~\bibnamefont
  {Albert}}\ and\ \bibinfo {author} {\bibfnamefont {A.-L.}\ \bibnamefont
  {Barab{\'a}si}},\ }\href@noop {} {\bibfield  {journal} {\bibinfo  {journal}
  {Reviews of modern physics}\ }\textbf {\bibinfo {volume} {74}},\ \bibinfo
  {pages} {47} (\bibinfo {year} {2002})}\BibitemShut {NoStop}%
\bibitem [{\citenamefont {Newman}(2010)}]{newmanbook}%
  \BibitemOpen
  \bibfield  {author} {\bibinfo {author} {\bibfnamefont {M.~E.}\ \bibnamefont
  {Newman}},\ }\href@noop {} {\emph {\bibinfo {title} {Networks: An
  Introduction}}}\ (\bibinfo  {publisher} {Oxford University Press},\ \bibinfo
  {address} {Oxford},\ \bibinfo {year} {2010})\BibitemShut {NoStop}%
\bibitem [{\citenamefont {Pikovsky}\ \emph {et~al.}(2001)\citenamefont
  {Pikovsky}, \citenamefont {Rosenblum},\ and\ \citenamefont
  {Kurths}}]{Pikovsky2001}%
  \BibitemOpen
  \bibfield  {author} {\bibinfo {author} {\bibfnamefont {A.}~\bibnamefont
  {Pikovsky}}, \bibinfo {author} {\bibfnamefont {M.}~\bibnamefont {Rosenblum}},
  \ and\ \bibinfo {author} {\bibfnamefont {J.}~\bibnamefont {Kurths}},\
  }\href@noop {} {\emph {\bibinfo {title} {Synchronization}}}\ (\bibinfo
  {publisher} {Cambridge University Press, Cambridge, UK},\ \bibinfo {year}
  {2001})\BibitemShut {NoStop}%
\bibitem [{\citenamefont {Arenas}\ \emph {et~al.}(2008)\citenamefont {Arenas},
  \citenamefont {D{\'\i}az-Guilera}, \citenamefont {Kurths}, \citenamefont
  {Moreno},\ and\ \citenamefont {Zhou}}]{arenasreview}%
  \BibitemOpen
  \bibfield  {author} {\bibinfo {author} {\bibfnamefont {A.}~\bibnamefont
  {Arenas}}, \bibinfo {author} {\bibfnamefont {A.}~\bibnamefont
  {D{\'\i}az-Guilera}}, \bibinfo {author} {\bibfnamefont {J.}~\bibnamefont
  {Kurths}}, \bibinfo {author} {\bibfnamefont {Y.}~\bibnamefont {Moreno}}, \
  and\ \bibinfo {author} {\bibfnamefont {C.}~\bibnamefont {Zhou}},\ }\href@noop
  {} {\bibfield  {journal} {\bibinfo  {journal} {Physics reports}\ }\textbf
  {\bibinfo {volume} {469}},\ \bibinfo {pages} {93} (\bibinfo {year}
  {2008})}\BibitemShut {NoStop}%
\bibitem [{\citenamefont {Asllani}\ \emph {et~al.}(2018)\citenamefont
  {Asllani}, \citenamefont {Expert},\ and\ \citenamefont
  {Carletti}}]{AECPlos2018}%
  \BibitemOpen
  \bibfield  {author} {\bibinfo {author} {\bibfnamefont {M.}~\bibnamefont
  {Asllani}}, \bibinfo {author} {\bibfnamefont {P.}~\bibnamefont {Expert}}, \
  and\ \bibinfo {author} {\bibfnamefont {T.}~\bibnamefont {Carletti}},\
  }\href@noop {} {\bibfield  {journal} {\bibinfo  {journal} {PLoS Comput Biol}\
  }\textbf {\bibinfo {volume} {14}},\ \bibinfo {pages} {e1006296} (\bibinfo
  {year} {2018})}\BibitemShut {NoStop}%
\bibitem [{\citenamefont {Motter}\ \emph {et~al.}(2013)\citenamefont {Motter},
  \citenamefont {Myers}, \citenamefont {Anghel},\ and\ \citenamefont
  {Nishikawa}}]{MMAN2013}%
  \BibitemOpen
  \bibfield  {author} {\bibinfo {author} {\bibfnamefont {A.~E.}\ \bibnamefont
  {Motter}}, \bibinfo {author} {\bibfnamefont {S.~A.}\ \bibnamefont {Myers}},
  \bibinfo {author} {\bibfnamefont {M.}~\bibnamefont {Anghel}}, \ and\ \bibinfo
  {author} {\bibfnamefont {T.}~\bibnamefont {Nishikawa}},\ }\href@noop {}
  {\bibfield  {journal} {\bibinfo  {journal} {Nature Physics}\ }\textbf
  {\bibinfo {volume} {9}},\ \bibinfo {pages} {191} (\bibinfo {year}
  {2013})}\BibitemShut {NoStop}%
\bibitem [{\citenamefont {Turing}(1952)}]{Turing}%
  \BibitemOpen
  \bibfield  {author} {\bibinfo {author} {\bibfnamefont {A.~M.}\ \bibnamefont
  {Turing}},\ }\href@noop {} {\bibfield  {journal} {\bibinfo  {journal} {Phil.
  Trans. R. Soc. Lond. B}\ }\textbf {\bibinfo {volume} {237}},\ \bibinfo
  {pages} {37} (\bibinfo {year} {1952})}\BibitemShut {NoStop}%
\bibitem [{\citenamefont {Nakao}\ and\ \citenamefont
  {Mikhailov}(2010)}]{NM2010}%
  \BibitemOpen
  \bibfield  {author} {\bibinfo {author} {\bibfnamefont {H.}~\bibnamefont
  {Nakao}}\ and\ \bibinfo {author} {\bibfnamefont {A.~S.}\ \bibnamefont
  {Mikhailov}},\ }\href@noop {} {\bibfield  {journal} {\bibinfo  {journal}
  {Nature Physics}\ }\textbf {\bibinfo {volume} {6}},\ \bibinfo {pages} {544}
  (\bibinfo {year} {2010})}\BibitemShut {NoStop}%
\bibitem [{\citenamefont {Stilwell}\ \emph {et~al.}(2006)\citenamefont
  {Stilwell}, \citenamefont {Bollt},\ and\ \citenamefont
  {Roberson}}]{SBGR2006}%
  \BibitemOpen
  \bibfield  {author} {\bibinfo {author} {\bibfnamefont {D.~J.}\ \bibnamefont
  {Stilwell}}, \bibinfo {author} {\bibfnamefont {E.~M.}\ \bibnamefont {Bollt}},
  \ and\ \bibinfo {author} {\bibfnamefont {D.~G.}\ \bibnamefont {Roberson}},\
  }\href@noop {} {\bibfield  {journal} {\bibinfo  {journal} {SIAM J. Applied
  Dynamical Systems}\ }\textbf {\bibinfo {volume} {5}},\ \bibinfo {pages} {140}
  (\bibinfo {year} {2006})}\BibitemShut {NoStop}%
\bibitem [{\citenamefont {Petit}\ \emph {et~al.}(2017)\citenamefont {Petit},
  \citenamefont {Lauwens}, \citenamefont {Fanelli},\ and\ \citenamefont
  {Carletti}}]{PABFC2017}%
  \BibitemOpen
  \bibfield  {author} {\bibinfo {author} {\bibfnamefont {J.}~\bibnamefont
  {Petit}}, \bibinfo {author} {\bibfnamefont {B.}~\bibnamefont {Lauwens}},
  \bibinfo {author} {\bibfnamefont {D.}~\bibnamefont {Fanelli}}, \ and\
  \bibinfo {author} {\bibfnamefont {T.}~\bibnamefont {Carletti}},\ }\href@noop
  {} {\bibfield  {journal} {\bibinfo  {journal} {Phys. Rev. Letters}\ }\textbf
  {\bibinfo {volume} {119}},\ \bibinfo {pages} {148301} (\bibinfo {year}
  {2017})}\BibitemShut {NoStop}%
\bibitem [{\citenamefont {Lucas}\ \emph {et~al.}(2018)\citenamefont {Lucas},
  \citenamefont {Fanelli}, \citenamefont {Carletti},\ and\ \citenamefont
  {Petit}}]{LFCP2018}%
  \BibitemOpen
  \bibfield  {author} {\bibinfo {author} {\bibfnamefont {M.}~\bibnamefont
  {Lucas}}, \bibinfo {author} {\bibfnamefont {D.}~\bibnamefont {Fanelli}},
  \bibinfo {author} {\bibfnamefont {T.}~\bibnamefont {Carletti}}, \ and\
  \bibinfo {author} {\bibfnamefont {J.}~\bibnamefont {Petit}},\ }\href@noop {}
  {\bibfield  {journal} {\bibinfo  {journal} {EPL}\ }\textbf {\bibinfo {volume}
  {121}},\ \bibinfo {pages} {50008} (\bibinfo {year} {2018})}\BibitemShut
  {NoStop}%
\bibitem [{\citenamefont {Amritkara}\ and\ \citenamefont
  {Hu}(2006)}]{AmritkarHu2006}%
  \BibitemOpen
  \bibfield  {author} {\bibinfo {author} {\bibfnamefont {R.}~\bibnamefont
  {Amritkara}}\ and\ \bibinfo {author} {\bibfnamefont {C.-K.}\ \bibnamefont
  {Hu}},\ }\href@noop {} {\bibfield  {journal} {\bibinfo  {journal} {Chaos}\
  }\textbf {\bibinfo {volume} {16}},\ \bibinfo {pages} {015117} (\bibinfo
  {year} {2006})}\BibitemShut {NoStop}%
\bibitem [{\citenamefont {Boccaletti}\ \emph {et~al.}(2006)\citenamefont
  {Boccaletti}, \citenamefont {Hwang}, \citenamefont {Chavez}, \citenamefont
  {Amann}, \citenamefont {J},\ and\ \citenamefont {Pecora}}]{BHCAKP}%
  \BibitemOpen
  \bibfield  {author} {\bibinfo {author} {\bibfnamefont {S.}~\bibnamefont
  {Boccaletti}}, \bibinfo {author} {\bibfnamefont {D.-W.}\ \bibnamefont
  {Hwang}}, \bibinfo {author} {\bibfnamefont {M.}~\bibnamefont {Chavez}},
  \bibinfo {author} {\bibfnamefont {A.}~\bibnamefont {Amann}}, \bibinfo
  {author} {\bibfnamefont {K.}~\bibnamefont {J}}, \ and\ \bibinfo {author}
  {\bibfnamefont {L.~M.}\ \bibnamefont {Pecora}},\ }\href@noop {} {\bibfield
  {journal} {\bibinfo  {journal} {Phys. Rev. E}\ }\textbf {\bibinfo {volume}
  {74}},\ \bibinfo {pages} {016102} (\bibinfo {year} {2006})}\BibitemShut
  {NoStop}%
\bibitem [{\citenamefont {Zhang}\ and\ \citenamefont
  {Strogatz}(2021)}]{ZS2021}%
  \BibitemOpen
  \bibfield  {author} {\bibinfo {author} {\bibfnamefont {Y.}~\bibnamefont
  {Zhang}}\ and\ \bibinfo {author} {\bibfnamefont {S.~H.}\ \bibnamefont
  {Strogatz}},\ }\href@noop {} {\bibfield  {journal} {\bibinfo  {journal}
  {Nature Communications}\ ,\ \bibinfo {pages} {3273}} (\bibinfo {year}
  {2021})}\BibitemShut {NoStop}%
\bibitem [{\citenamefont {Van~Gorder}(2021{\natexlab{a}})}]{vangorder2}%
  \BibitemOpen
  \bibfield  {author} {\bibinfo {author} {\bibfnamefont {R.~A.}\ \bibnamefont
  {Van~Gorder}},\ }\href@noop {} {\bibfield  {journal} {\bibinfo  {journal}
  {Proceedings of the Royal Society A}\ }\textbf {\bibinfo {volume} {477}},\
  \bibinfo {pages} {20200753} (\bibinfo {year}
  {2021}{\natexlab{a}})}\BibitemShut {NoStop}%
\bibitem [{\citenamefont {Pecora}\ and\ \citenamefont
  {Carroll}(1998)}]{Pecora}%
  \BibitemOpen
  \bibfield  {author} {\bibinfo {author} {\bibfnamefont {L.~M.}\ \bibnamefont
  {Pecora}}\ and\ \bibinfo {author} {\bibfnamefont {T.~L.}\ \bibnamefont
  {Carroll}},\ }\href@noop {} {\bibfield  {journal} {\bibinfo  {journal} {Phys.
  Rev. Letters}\ }\textbf {\bibinfo {volume} {80}},\ \bibinfo {pages} {2109}
  (\bibinfo {year} {1998})}\BibitemShut {NoStop}%
\bibitem [{\citenamefont {Holme}\ and\ \citenamefont
  {Saram{\"a}ki}(2013)}]{Holme2013}%
  \BibitemOpen
  \bibfield  {author} {\bibinfo {author} {\bibfnamefont {P.}~\bibnamefont
  {Holme}}\ and\ \bibinfo {author} {\bibfnamefont {J.}~\bibnamefont
  {Saram{\"a}ki}},\ }\href@noop {} {\emph {\bibinfo {title} {Temporal
  Networks}}}\ (\bibinfo  {publisher} {Springer-Verlag, Berlin},\ \bibinfo
  {year} {2013})\BibitemShut {NoStop}%
\bibitem [{\citenamefont {Masuda}\ and\ \citenamefont
  {Lambiotte}(2016)}]{MR2016}%
  \BibitemOpen
  \bibfield  {author} {\bibinfo {author} {\bibfnamefont {N.}~\bibnamefont
  {Masuda}}\ and\ \bibinfo {author} {\bibfnamefont {R.}~\bibnamefont
  {Lambiotte}},\ }\href@noop {} {\emph {\bibinfo {title} {A Guide to Temporal
  Networks}}}\ (\bibinfo  {publisher} {World Scientific, London},\ \bibinfo
  {year} {2016})\BibitemShut {NoStop}%
\bibitem [{\citenamefont {Kuramoto}(1984)}]{Kuramoto}%
  \BibitemOpen
  \bibfield  {author} {\bibinfo {author} {\bibfnamefont {Y.}~\bibnamefont
  {Kuramoto}},\ }\href@noop {} {\emph {\bibinfo {title} {Chemical oscillations,
  waves, and turbulence}}}\ (\bibinfo  {publisher} {Springer-Verlag, New
  York},\ \bibinfo {year} {1984})\BibitemShut {NoStop}%
\bibitem [{\citenamefont {Ghosh}\ \emph {et~al.}(2021)\citenamefont {Ghosh},
  \citenamefont {Frasca}, \citenamefont {Rizzo}, \citenamefont {Majhi},
  \citenamefont {Rakshit}, \citenamefont {Alfaro-Bittner},\ and\ \citenamefont
  {Boccaletti}}]{GFRMRABB2022}%
  \BibitemOpen
  \bibfield  {author} {\bibinfo {author} {\bibfnamefont {D.}~\bibnamefont
  {Ghosh}}, \bibinfo {author} {\bibfnamefont {M.}~\bibnamefont {Frasca}},
  \bibinfo {author} {\bibfnamefont {A.}~\bibnamefont {Rizzo}}, \bibinfo
  {author} {\bibfnamefont {S.}~\bibnamefont {Majhi}}, \bibinfo {author}
  {\bibfnamefont {S.}~\bibnamefont {Rakshit}}, \bibinfo {author} {\bibfnamefont
  {K.}~\bibnamefont {Alfaro-Bittner}}, \ and\ \bibinfo {author} {\bibfnamefont
  {S.}~\bibnamefont {Boccaletti}},\ }\href@noop {} {\bibfield  {journal}
  {\bibinfo  {journal} {arXiv preprint arXiv:2109.07618}\ } (\bibinfo {year}
  {2021})}\BibitemShut {NoStop}%
\bibitem [{\citenamefont {Huang}\ \emph {et~al.}(2009)\citenamefont {Huang},
  \citenamefont {Chen}, \citenamefont {Lai},\ and\ \citenamefont
  {Pecora}}]{HCLP}%
  \BibitemOpen
  \bibfield  {author} {\bibinfo {author} {\bibfnamefont {L.}~\bibnamefont
  {Huang}}, \bibinfo {author} {\bibfnamefont {Q.}~\bibnamefont {Chen}},
  \bibinfo {author} {\bibfnamefont {Y.-C.}\ \bibnamefont {Lai}}, \ and\
  \bibinfo {author} {\bibfnamefont {L.~M.}\ \bibnamefont {Pecora}},\
  }\href@noop {} {\bibfield  {journal} {\bibinfo  {journal} {Phys. Rev. E}\
  }\textbf {\bibinfo {volume} {80}},\ \bibinfo {pages} {036204} (\bibinfo
  {year} {2009})}\BibitemShut {NoStop}%
\bibitem [{\citenamefont {Van~Gorder}(2021{\natexlab{b}})}]{vangorder3}%
  \BibitemOpen
  \bibfield  {author} {\bibinfo {author} {\bibfnamefont {R.~A.}\ \bibnamefont
  {Van~Gorder}},\ }\href@noop {} {\bibfield  {journal} {\bibinfo  {journal}
  {Phil. Trans. A}\ }\textbf {\bibinfo {volume} {379}},\ \bibinfo {pages}
  {20210001} (\bibinfo {year} {2021}{\natexlab{b}})}\BibitemShut {NoStop}%
\bibitem [{\citenamefont {van Harten}(1991)}]{vanharten}%
  \BibitemOpen
  \bibfield  {author} {\bibinfo {author} {\bibfnamefont {A.}~\bibnamefont {van
  Harten}},\ }\href@noop {} {\bibfield  {journal} {\bibinfo  {journal} {J.
  Nonlinear Sci.}\ }\textbf {\bibinfo {volume} {1}},\ \bibinfo {pages} {397}
  (\bibinfo {year} {1991})}\BibitemShut {NoStop}%
\bibitem [{\citenamefont {Aranson}\ and\ \citenamefont
  {Kramer}(2002)}]{aranson}%
  \BibitemOpen
  \bibfield  {author} {\bibinfo {author} {\bibfnamefont {I.}~\bibnamefont
  {Aranson}}\ and\ \bibinfo {author} {\bibfnamefont {L.}~\bibnamefont
  {Kramer}},\ }\href@noop {} {\bibfield  {journal} {\bibinfo  {journal}
  {Reviews of Modern Physics}\ }\textbf {\bibinfo {volume} {74}},\ \bibinfo
  {pages} {99} (\bibinfo {year} {2002})}\BibitemShut {NoStop}%
\bibitem [{\citenamefont {Garca-Morales}\ and\ \citenamefont
  {Krischer}(2012)}]{garcamorales}%
  \BibitemOpen
  \bibfield  {author} {\bibinfo {author} {\bibfnamefont {V.}~\bibnamefont
  {Garca-Morales}}\ and\ \bibinfo {author} {\bibfnamefont {K.}~\bibnamefont
  {Krischer}},\ }\href@noop {} {\bibfield  {journal} {\bibinfo  {journal}
  {Contem. Phys.}\ }\textbf {\bibinfo {volume} {53}},\ \bibinfo {pages} {79}
  (\bibinfo {year} {2012})}\BibitemShut {NoStop}%
\bibitem [{\citenamefont {MATLAB}(2021)}]{MATLAB2021}%
  \BibitemOpen
  \bibfield  {author} {\bibinfo {author} {\bibnamefont {MATLAB}},\ }\href@noop
  {} {\emph {\bibinfo {title} {Version: 9.10.0.1602886 (R2021a)}}}\ (\bibinfo
  {publisher} {The MathWorks Inc.},\ \bibinfo {address} {Natick,
  Massachusetts},\ \bibinfo {year} {2021})\BibitemShut {NoStop}%
\bibitem [{\citenamefont {Perron}(1930)}]{Perron1930}%
  \BibitemOpen
  \bibfield  {author} {\bibinfo {author} {\bibfnamefont {O.}~\bibnamefont
  {Perron}},\ }\href@noop {} {\bibfield  {journal} {\bibinfo  {journal}
  {Mathematische Zeitschrift}\ }\textbf {\bibinfo {volume} {32}},\ \bibinfo
  {pages} {703} (\bibinfo {year} {1930})}\BibitemShut {NoStop}%
\bibitem [{\citenamefont {Vinograd}(1952)}]{Vinograd1952}%
  \BibitemOpen
  \bibfield  {author} {\bibinfo {author} {\bibfnamefont {R.}~\bibnamefont
  {Vinograd}},\ }\href@noop {} {\bibfield  {journal} {\bibinfo  {journal}
  {Doklady Akad. Nauk SSSR}\ }\textbf {\bibinfo {volume} {84}},\ \bibinfo
  {pages} {201} (\bibinfo {year} {1952})}\BibitemShut {NoStop}%
\bibitem [{\citenamefont {Lorenz}(1963)}]{Lorenz1963}%
  \BibitemOpen
  \bibfield  {author} {\bibinfo {author} {\bibfnamefont {E.~N.}\ \bibnamefont
  {Lorenz}},\ }\href@noop {} {\bibfield  {journal} {\bibinfo  {journal} {J.
  Atmos. Sci.}\ }\textbf {\bibinfo {volume} {20}},\ \bibinfo {pages} {130}
  (\bibinfo {year} {1963})}\BibitemShut {NoStop}%
\end{thebibliography}%

\appendix
\section{About the small network used in the main text}
\label{sec:smallnet}
In the main text we decided to test our theory by using a small time dependent network, whose eigenvalues and eigenvectors exhibit a simple behaviour. In this way we aimed at removing any unnecessary complications and focusing our attention on the crucial role played by matrix $\mathbf{c}(t)$. Moreover, the example was constructed in such a way to yield  a constant $\mathbf{c}$, allowing to analytically solve Eq.~\eqref{eq:GLHGlinalpha3compact}.

Let us thus fix $n=3$, a real positive $\Omega$ and define 
\begin{equation*}
 \mathbf{c}=\left(
\begin{matrix}
 0 & 0 & 0\\
 0 & 0 & \Omega\\
 0 & -\Omega& 0
\end{matrix}\right)\, .
\end{equation*}
Our goal is to determine the network adjacency matrix and for this we need to compute the time evolution of the eigenvectors. By Eq.~\eqref{eq:cab}, we have $ \frac{d\vec{\phi}^{(\beta)}}{dt}=\sum_\alpha c_{\beta\alpha}\vec{\phi}^{(\alpha)}$, that is 
\begin{equation*}
\begin{dcases}
 \frac{d\vec{\phi}^{(1)}}{dt}=0\\ 
 \frac{d\vec{\phi}^{(2)}}{dt}=\sum_\alpha c_{2\alpha}\vec{\phi}^{(\alpha)}=\Omega\vec{\phi}^{(3)}\\ 
 \frac{d\vec{\phi}^{(3)}}{dt}=\sum_\alpha c_{3\alpha}\vec{\phi}^{(\alpha)}=-\Omega\vec{\phi}^{(2)}\, .
\end{dcases}
\end{equation*}
The eigenvector associated to $\Lambda^{(1)}=0$ is constant and  given by $\vec{\phi}^{(1)}=(1,1,1)^\top/\sqrt{3}$. The two other eigenvectors are solutions of
\begin{equation*}
\vec{\phi}^{(2)}(t)=\vec{a}\cos(\Omega t)+\vec{b}\sin(\Omega t)\text{ and } \vec{\phi}^{(3)}(t)=-\vec{a}\sin(\Omega t)+\vec{b}\cos(\Omega t)\, .
\end{equation*}
The unknown vectors $\vec{a}$ and $\vec{b}$ should be determined by using the initial conditions for the eigenvectors, $\vec{\phi}^{(2)}(0)=\vec{a}$ and $\vec{\phi}^{(3)}(0)=\vec{b}$. Moreover they should satisfy the following constraints to have an orthonormal basis for all $t$
\begin{equation*}
 |\vec{a}|=|\vec{b}|=1\, , \vec{a}^\top\cdot \vec{b}=0\text{ and }\vec{a}^\top\cdot \phi^{(1)}=\vec{b}^\top\cdot \phi^{(1)}=0\, ,
\end{equation*}
for a sake of definitiveness let us take
\begin{equation*}
 \vec{a}=\frac{1}{\sqrt{6}}\left(
\begin{matrix}
 1\\-2\\1
\end{matrix}\right)\text{ and } \vec{b}=\frac{1}{\sqrt{2}}\left(
\begin{matrix}
 -1\\0\\1
\end{matrix}\right)\, .
\end{equation*}
Finally the eigenvectors are given by
\begin{eqnarray}
\label{eq:eigex}
 \vec{\phi}^{(2)}(t)&=&\frac{1}{\sqrt{6}}\left(
\begin{smallmatrix}
 1\\-2\\1
\end{smallmatrix}\right)\cos(\Omega t)+\frac{1}{\sqrt{2}}\left(
\begin{smallmatrix}
 -1\\0\\1
\end{smallmatrix}\right)\sin(\Omega t)\notag\text{ and }\\  \vec{\phi}^{(3)}(t)&=&-\frac{1}{\sqrt{6}}\left(
\begin{smallmatrix}
 1\\-2\\1
\end{smallmatrix}\right)\sin(\Omega t)+\frac{1}{\sqrt{2}}\left(
\begin{smallmatrix}
 -1\\0\\1
\end{smallmatrix}\right)\cos(\Omega t)\, .
\end{eqnarray}

\begin{figure*}[ht]
\centering
\includegraphics[scale=0.18]{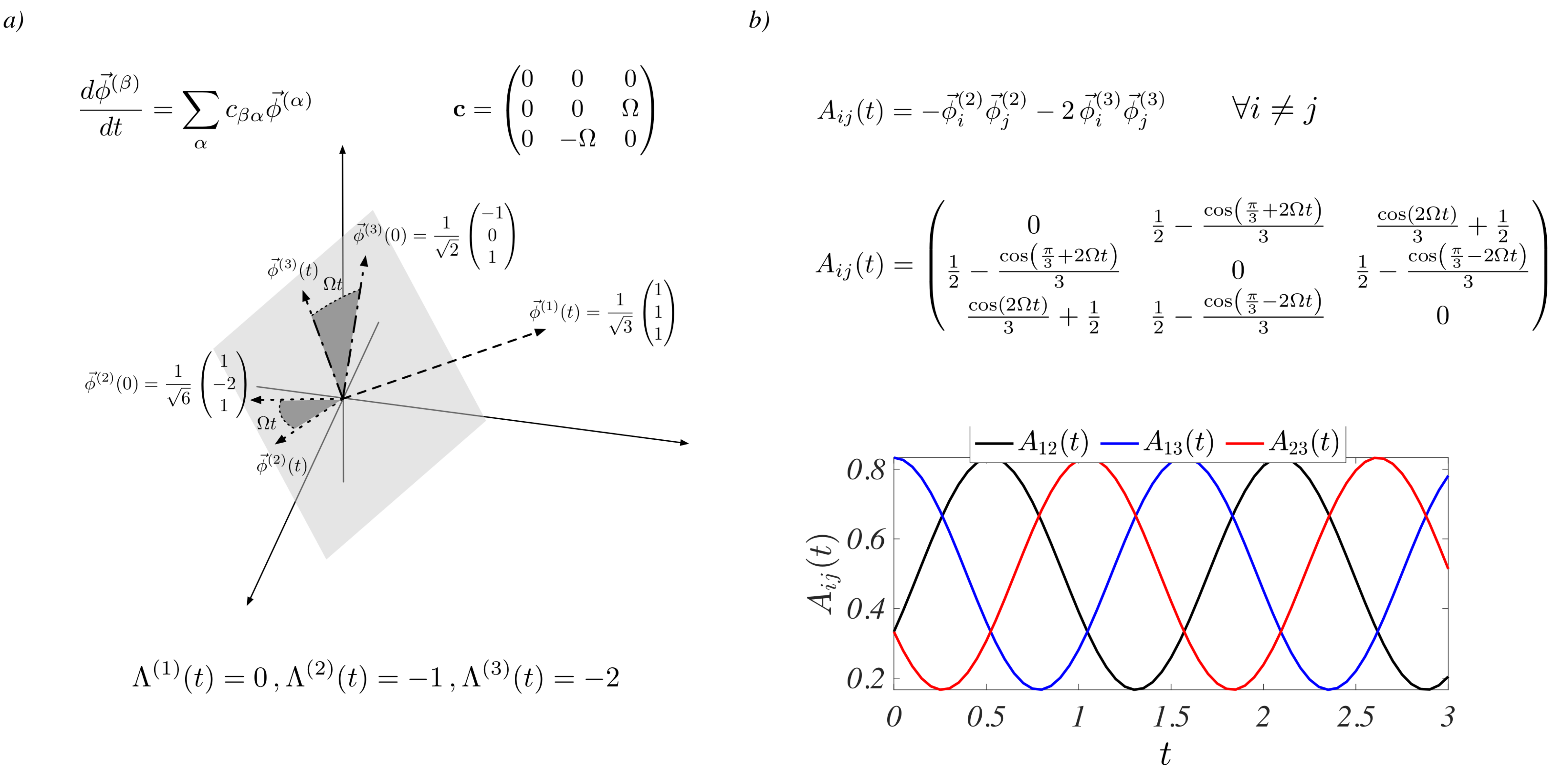}
%\vspace{-1.5cm}
\caption{\textbf{Construction of the simple network.} In panel a) we show the time evolution of the eigenvectors ruled by Eq.~\eqref{eq:cab}. For a given choice of the matrix $\mathbf{c}$ and of the initial eigenbasis, we schematically represents the dynamics of the latter: $\vec{\phi}^{(2)}(t)$ and $\vec{\phi}^{(3)}(t)$ rotate by an angle $\Omega t$ while laying on the plane (grey) orthogonal to $\vec{\phi}^{(1)}$. For a given choice of the eigenvalues, we report in panel b) the analytical expression of the adjacency matrix $\mathbf{A}$ and the time evolution of its entries.}
\label{fig:simplenet}
\end{figure*}

To compute the Laplace matrix, let us fix the remaining two eigenvalues $\Lambda^{(2)}=-1$ and $\Lambda^{(3)}=-2$, then recalling that $ L_{ij}(t)=\sum_\alpha \Lambda^{(\alpha)}\phi^{(\alpha)}_i(t)\phi^{(\alpha)}_j(t)$ we have
\begin{equation*}
 L_{ij}(t) = -\phi^{(2)}_i(t)\phi^{(2)}_j(t)-2\phi^{(3)}_i(t)\phi^{(3)}_j(t)\, .
\end{equation*}
Finally, $A_{ij}(t) = L_{ij}(t)$ for all $i\neq j$ and $A_{ii}(t)=0$,
%:
%\begin{equation}
% A_{ij}(t)=\left(\begin{matrix} 0 & \frac{1}{2}-\frac{\cos\left(\frac{\pi }{3}+2\Omega t\right)}{3} & \frac{\cos\left(2\Omega t\right)}{3}+\frac{1}{2}\\ \frac{1}{2}-\frac{\cos\left(\frac{\pi }{3}+2\Omega t\right)}{3} & 0 & \frac{1}{2}-\frac{\cos\left(\frac{\pi }{3}-2\Omega t\right)}{3}\\ \frac{\cos\left(2\Omega t\right)}{3}+\frac{1}{2} & \frac{1}{2}-\frac{\cos\left(\frac{\pi }{3}-2\Omega t\right)}{3} & 0 \end{matrix}\right)\, , 
%\label{eq:adjt}
%\end{equation}
whose explicit formula and time evolution are reported in panel b) of Fig.~\ref{fig:simplenet}~\footnote{To better appreciate the dynamics of the network, we provide in the accompanying Supplementary Movie $1$, a graphical evolution of the network, the value of $A_{ij}(t)$ and the position of the eigenvectors in space. The latter ones appear to rotate with constant angular velocity $\Omega$ in the plane orthogonal to the eigenvector $\phi^{(1)}$, as it should being $e^{\mathbf{c}t}$ an orthogonal matrix.}. We can observe that for all $i\neq j$, $A_{ij}(t)>0$ for all $t$, we thus have a connected network whose links have weights that oscillate in time.

\begin{remark}[About the derivative of the Laplace matrix]
\label{rem:dLdt}
 Given the adjacency matrix of the time varying network, we can compute its associated time varying Laplace matrix that can be decomposed into the eigenbasis as : 
 \begin{equation}
\label{eq:Leigdec}
L_{ij}=\sum_\alpha \Lambda^{(\alpha)}\phi^{(\alpha)}_i\phi^{(\alpha)}_j\, .
\end{equation}
We are interested in relating the time derivative of $\mathbf{L}$ to known quantities, i.e., $\Lambda^{(\alpha)}(t)$ and $\phi^{(\alpha)}(t)$.
 
 This can be done by observing that one can write $\frac{dL_{ij}}{dt}=\sum_{\alpha \beta}\ell_{\alpha\beta}(t)\phi^{(\alpha)}_i\phi^{(\beta)}_j$ for suitable coefficients $\ell_{\alpha\beta}(t)$, that is the projections of the matrix $\frac{dL_{ij}}{dt}$ onto the matrix basis $\phi^{(\alpha)}_i\phi^{(\beta)}_j$. From the definition
\begin{equation*}
 \frac{dL_{ij}}{dt}=\sum_\alpha \frac{d\Lambda^{(\alpha)}}{dt}\phi^{(\alpha)}_i\phi^{(\alpha)}_j+\sum_\alpha \Lambda^{(\alpha)}\frac{d\phi^{(\alpha)}_i}{dt}\phi^{(\alpha)}_j+\sum_\alpha \Lambda^{(\alpha)}\phi^{(\alpha)}_i\frac{d\phi^{(\alpha)}_j}{dt}\, ,
\end{equation*}
and thus
\begin{equation*}
 \ell_{\alpha\beta}=\sum_{ij}\phi^{(\alpha)}_i\frac{dL_{ij}}{dt}\phi^{(\beta)}_j=\frac{d\Lambda^{(\alpha)}}{dt}\delta_{\alpha\beta}+\left(\Lambda^{(\alpha)}-\Lambda^{(\beta)}\right) \sum_i \frac{d\phi^{(\alpha)}}{dt}\phi_j^{(\beta)}=\frac{d\Lambda^{(\alpha)}}{dt}\delta_{\alpha\beta}+\left(\Lambda^{(\alpha)}-\Lambda^{(\beta)}\right) c_{\alpha\beta}\, .
\end{equation*}
In conclusion
\begin{equation}
\label{eq:ellalphabeta}
 \ell_{\alpha\alpha}=\frac{d\Lambda^{(\alpha)}}{dt}\quad\forall \alpha\text{ and }\ell_{\alpha\beta}=\left(\Lambda^{(\alpha)}-\Lambda^{(\beta)}\right) c_{\alpha\beta}\quad\forall \beta\neq \alpha\, .
\end{equation}
\end{remark}

\begin{remark}[About the periodicity]
\label{rem:periodicity}
Let us assume the network varies periodically in time with some period $T>0$. We are interested in determining the conditions for which  the spectrum, i.e., $\Lambda^{(\alpha)}(t)$ and $\phi^{(\alpha)}(t)$, also exhibits the same behaviour.

Being the network $T$-periodic means that the adjacency matrix is $T$-periodic and so does the Laplace matrix and its derivative. We can then use the first relation of~\eqref{eq:ellalphabeta} to write
\begin{eqnarray*}
 \Lambda^{(\alpha)}(t+T)- \Lambda^{(\alpha)}(0)&=&\int_0^{t+T}\frac{d\Lambda^{(\alpha)}}{dt}(s) \,ds=\int_0^{t+T}\ell_{\alpha\alpha}(s) \,ds=\int_0^{t}\ell_{\alpha\alpha}(s) \,ds+\int_t^{t+T}\ell_{\alpha\alpha}(s) \,ds\\
&=& \Lambda^{(\alpha)}(t)- \Lambda^{(\alpha)}(0) +\int_t^{t+T}\ell_{\alpha\alpha}(s) \,ds=\Lambda^{(\alpha)}(t)- \Lambda^{(\alpha)}(0) +T \langle\ell_{\alpha\alpha} \rangle\, ,
\end{eqnarray*}
where in the last step we used the periodicity of $\ell_{\alpha\alpha}$ to rewrite the integral as the time average of the function. We can then conclude that $ \Lambda^{(\alpha)}$ is $T$-periodic if and only if $\langle\ell_{\alpha\alpha} \rangle=0$. 

Let us now consider the eigenvectors, then by definition
\begin{equation*}
 L(t+T)\phi^{(\alpha)}(t+T) =  \Lambda^{(\alpha)}(t+T)\phi^{(\alpha)}(t+T)\, ,
\end{equation*}
and by recalling the periodicity of $\mathbf{L}$ and $\Lambda^{(\alpha)}$ we can also write
\begin{equation*}
 L(t)\phi^{(\alpha)}(t+T) =  \Lambda^{(\alpha)}(t)\phi^{(\alpha)}(t+T)\, .
\end{equation*}
Namely $\phi^{(\alpha)}(t+T)$ satisfies the same eigenvector equation than $\phi^{(\alpha)}(t)$ and by the uniqueness of the eigenvectors we conclude that $\phi^{(\alpha)}(t+T)=\phi^{(\alpha)}(t)$, that is the eigenvector is also $T$-periodic.
\end{remark}

To conclude this section let us briefly introduce a generalisation of the simple network shown above allowing us to deal with the possibility to ``suddenly'' add or remove a link. For sake of simplicity we still consider the network to be made by three nodes. We will thus construct a network whose adjacency matrix at time $t=0$ is given by
\begin{equation}
\mathbf{A}(0)=\left(\begin{matrix}
 0& 1 &0\\
 1 & 0 & 1\\
 0 & 1 &0
\end{matrix}
\right)
\label{eq:Adjadd}
\end{equation}
namely node $2$ is connected to both node $1$ and $3$, that are instead disconnected. Then periodically, the latter nodes are linked together during a give time interval and then disconnected again.

More precisely, let us thus consider a time dependent matrix $\mathbf{c}(t)$ given by:
\begin{equation*}
 \mathbf{c}=\psi^\prime(t)\left(
\begin{matrix}
 0 & 0 & 0\\
 0 & 0 & 1\\
 0 & -1& 0
\end{matrix}\right)\, .
\end{equation*}
where $\psi^\prime$ is the derivative of a given periodic function hereafter specified, such that $\psi(0)=0$. We are interested in solving Eq.~\eqref{eq:cab}
\begin{equation*}
\begin{dcases}
 \frac{d\vec{\phi}^{(1)}}{dt}=0\\ 
 \frac{d\vec{\phi}^{(2)}}{dt}=\psi^\prime(t)\vec{\phi}^{(3)}\\ 
 \frac{d\vec{\phi}^{(3)}}{dt}=-\psi^\prime(t)\vec{\phi}^{(2)}\, ,
\end{dcases}
\end{equation*}
 for the following initial eigenbasis resulting from the choice of the adjacency matrix~\eqref{eq:Adjadd} at time $t=0$:
 \begin{equation*}
 \vec{\phi}^{(1)}(0)=\frac{1}{\sqrt{3}}\left(\begin{smallmatrix}1\\1\\1\end{smallmatrix}\right)\, , \vec{\phi}^{(2)}(0)=\frac{1}{\sqrt{2}}\left(\begin{smallmatrix}-1\\0\\1\end{smallmatrix}\right)\, , \vec{\phi}^{(3)}(0)=\frac{1}{\sqrt{6}}\left(\begin{smallmatrix}-1\\2\\-1\end{smallmatrix}\right)\, .
\end{equation*}
The associated eigenvalues are fixed to the values $\Lambda^{(1)}=0$, $\Lambda^{(2)}=-1$ and $\Lambda^{(3)}=-3$. Once again, to focus on the impact of the eigenvectors evolution, we set stationary eigenvalues, so to disentangle the two possible factors.

A straightforward computation allows to determine 
\begin{eqnarray}
\label{eq:eigex2}
 \vec{\phi}^{(2)}(t)&=&\frac{1}{\sqrt{6}}\left(
\begin{smallmatrix}
 1\\-2\\1
\end{smallmatrix}\right)\cos(\psi(t))+\frac{1}{\sqrt{2}}\left(
\begin{smallmatrix}
 -1\\0\\1
\end{smallmatrix}\right)\sin(\psi(t))\text{ and}\notag\\
  \vec{\phi}^{(3)}(t)&=&-\frac{1}{\sqrt{6}}\left(
\begin{smallmatrix}
 1\\-2\\1
\end{smallmatrix}\right)\sin(\psi(t))+\frac{1}{\sqrt{2}}\left(
\begin{smallmatrix}
 -1\\0\\1
\end{smallmatrix}\right)\cos(\psi(t))\, .
\end{eqnarray}

\begin{figure*}[ht]
\centering
\includegraphics[scale=0.24]{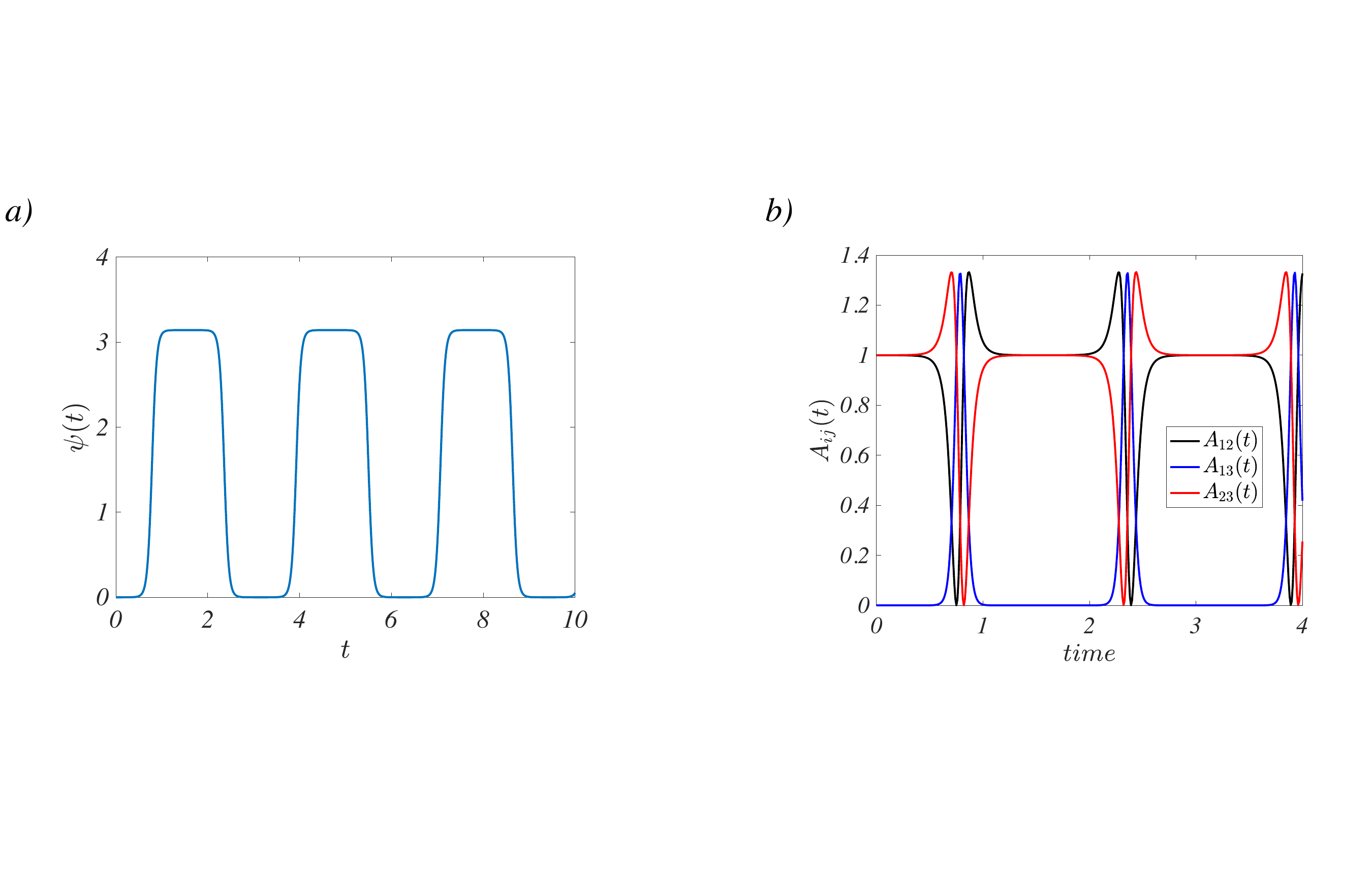}
\vspace{-1.5cm}
\caption{\textbf{Construction of a second simple network.} In the left panel a) we show the ``almost step'' function $\psi(t)$ used to create and to destroy a given link. In the right panel b) we report the time evolution of the entries of the computed adjacency matrix $\mathbf{A}(t)$.}
\label{fig:net2}
\end{figure*}

To reproduce the sought variation of the link $13$ one can use the following function
\begin{equation*}
 \psi(t)=\left[1+\tanh \left(M \left(\sin^2 t-1/2\right)\right)\right]\frac{\pi}{2}\, ,
\end{equation*}
whose graph is shown in the left panel of Fig.~\eqref{fig:net2}, for the value of the parameter $M=10$. By using the relation existing among the adjacency matrix and the Laplace matrix and the above obtained time varying eigenvectors, we can explicitly compute the entries of the adjacency matrix, whose time behaviour is reported in the right panel of Fig.~\eqref{fig:net2}. One can observe that for a relatively long period of time the link $13$ is not present, indeed $A_{13}(t)=0$ (blue line); then quite quickly the link $13$ is created and at the same time the two other links change of intensity reaching each one the null value for a single time instant (black and red lines). Finally the weight of the link $13$ decreases and reaches again the null value. This scheme is then periodically repeated in time~\footnote{To emphasise the sudden creation / removing dynamics, we provide in the accompanying Supplementary Movie $2$, a graphical evolution of the network, the value of $A_{ij}(t)$ and the position of the eigenvectors in space. The latter ones appear to rotate with non constant angular velocity $\psi(t)$ in the plane orthogonal to the eigenvector $\phi^{(1)}$, as it should being $e^{\int_0^t \mathbf{c}(s)ds}$ an orthogonal matrix.}.

\section{Master Stability Equation of time varying networks}
\label{sec:MSEtv}
The goal of this section is to provide to the interested reader the detailed computations allowing to derive the Master Stability Equation (MSE), Eq.~\eqref{eq:GLHGlinalpha3compact} in the main text.

Let us start by obtaining equation~\eqref{eq:maineq}, which rules the dynamics of the generic nonlinear system obtained by coupling together $n$ copies of a basic aspatial system $\frac{d\mathbf{x}}{dt}=\mathbf{F}(\mathbf{x})$. Observe that we here assume the number of nodes $n$ to be constant. This is not however a restrictive working hypothesis; indeed one can always assume to deal with a larger reservoir of nodes, disconnected from each other, and mimic the creation of a new node with the appearance of a link pointing to a node of the reservoir. 

Finally we assume  the coupling to vary in time and to exhibit a diffusive-like character, i.e., to depend on the difference of some nonlinear coupling function, $\mathbf{H}$, computed on adjacent nodes. In formula
\begin{equation*}
%\label{eq:maineqApp}
\frac{d\mathbf{x}_i}{dt}=\mathbf{F}(\mathbf{x}_i) +\varepsilon \sum_{j} A_{ij}(t)\left[ \mathbf{H}(\mathbf{x}_j)- \mathbf{H}(\mathbf{x}_i)\right]\, ,
\end{equation*}  
where $\mathbf{x}_i=(x_i^{(1)},\dots,x_i^{(d)})^\top$ represents the state of the $i$-th node, $\varepsilon>0$ is the strength of the coupling and $A_{ij}(t)$ are the entries of the (symmetric) adjacency matrix encoding for the underlying time dependent network. By recalling the definition of node degree, $k_i(t)=\sum_j A_{ij}(t)$, and of the Laplace matrix, $L_{ij}(t) =A_{ij}(t)-\delta_{ij}k_i(t)$, we can rewrite the previous equation as
\begin{equation}
\label{eq:maineq2App}
\frac{d\mathbf{x}_i}{dt}=\mathbf{F}(\mathbf{x}_i) +\varepsilon \sum_{j} L_{ij}(t) \mathbf{H}(\mathbf{x}_j)\, .
\end{equation}  
Let us observe that to keep the notation consistent throughout the work, we hereby define the Laplace matrix to be nonpositive defined at odd with the assumption elsewhere used in the literature. This choice yields the plus sign in front of the coupling term. Consider now a solution $\mathbf{s}(t)$ of the basic system, i.e., $\frac{d\mathbf{s}}{dt}=\mathbf{F}(\mathbf{s})$, and assume it to be stable. Recalling the zero-sum property of the Laplace matrix, i.e., $\sum_j L_{ij}(t)=0$ for all $i=1,\dots, n$ and all $t$, one can straightforwardly show that $\mathbf{s}(t)$ solves also Eq.~\eqref{eq:maineq2App}. Moreover, because of the stability assumption, one can prove that $\mathbf{s}(t)$ is also a stable solution for the coupled system with respect to homogeneous perturbation, which corresponds to simultaneously applying the same noisy small term to all the nodes. In this way the system is able to display a global synchronous behaviour.

To analyse the stability of the reference solution with respect to heterogeneous, i.e., node dependent, perturbations one can introduce the variables $\delta\mathbf{x}_i=\mathbf{x}_i-\mathbf{s}$, and, under the assumption of small displacements,  determine their time evolution at the linear order of approximation. We hence linearise Eq.~\eqref{eq:maineq2App} about the reference solution to obtain
\begin{equation}
\label{eq:GLHGlinApp}
\frac{d\delta\mathbf{x}_i}{dt}=\mathbf{J}_\mathbf{F}(\mathbf{s}(t))\delta\mathbf{x}_i +\varepsilon \sum_{j} {L}_{ij}(t)\mathbf{J}_\mathbf{H}(\mathbf{s}(t))\delta\mathbf{x}_j\, , 
\end{equation}
where $\mathbf{J}_\mathbf{F}(\mathbf{s}(t))$ is the Jacobian of the function $\mathbf{F}$ evaluated on the reference solution and, similarly, $\mathbf{J}_\mathbf{H}(\mathbf{s}(t))$ is the Jacobian of the function $\mathbf{H}$. In this way we obtained a linear non-autonomous (matrix) ODE, whose solution allows us to determine the asymptotic decay (or growth) of the heterogeneous perturbation and thus conclude on the stability of the homogenous state.

The information about the coupling is fully contained in the matrix $\mathbf{L}$. In the spirit of the works by Pecora and collaborators~\cite{Pecora,HCLP}, we can project the displacements $\delta\mathbf{x}_i$ onto the Laplace eigenbasis, $\vec{\phi}^{(\alpha)}$, $\alpha=1,\dots,n$, to decouple the contribution coming from each eigenmode. Inserting thus $\delta\mathbf{x}_i=\sum_\alpha \delta\hat{\mathbf{x}}_{\alpha}\phi^{(\alpha)}_i$ into Eq.~\eqref{eq:GLHGlinApp} we get
\begin{eqnarray}
\label{eq:GLHGlinalphaApp}
\sum_\alpha \frac{d\delta\hat{\mathbf{x}}_{\alpha}}{dt}\phi^{(\alpha)}_i + \sum_\alpha \delta\hat{\mathbf{x}}_{\alpha}\frac{d\phi^{(\alpha)}_i}{dt} &=&\sum_\alpha \mathbf{J}_\mathbf{F}(\mathbf{s})\delta\hat{\mathbf{x}}_{\alpha}\phi^{(\alpha)}_i+\varepsilon \sum_\alpha\sum_j L_{ij}\mathbf{J}_\mathbf{H}(\mathbf{s})\delta\hat{\mathbf{x}}_{\alpha}\phi^{(\alpha)}_j\notag\\
&=&\sum_\alpha \mathbf{J}_\mathbf{F}(\mathbf{s})\delta\hat{\mathbf{x}}_{\alpha}\phi^{(\alpha)}_i+\varepsilon \sum_\alpha \Lambda^{(\alpha)}\mathbf{J}_\mathbf{H}(\mathbf{s})\delta\hat{\mathbf{x}}_{\alpha}\phi^{(\alpha)}_i\, , 
\end{eqnarray}
where we used the eigenvector definition to write $\sum_j L_{ij}\phi^{(\alpha)}_j=\Lambda^{(\alpha)}\phi^{(\alpha)}_i$ and to lighten the notations we did not explicitly write the dependence on $t$ of the involved variables. We can further exploit the orthonormality condition, $\sum_i \phi^{(\alpha)}_i\phi^{(\beta)}_i=\delta_{\alpha\beta}$. Hence, by multiplying the latter equation by $\phi^{(\beta)}_i$ and summing over $i$, we get
\begin{equation}
\label{eq:GLHGlinalpha2App}
\frac{d\delta\hat{\mathbf{x}}_{\beta}}{dt}+ \sum_\alpha \sum_i\delta\hat{\mathbf{x}}_{\alpha}\phi^{(\beta)}_i\frac{d\phi^{(\alpha)}_i}{dt} = \mathbf{J}_\mathbf{F}(\mathbf{s})\delta\hat{\mathbf{x}}_{\beta}+\varepsilon \Lambda^{(\beta)}\mathbf{J}_\mathbf{H}(\mathbf{s})\delta\hat{\mathbf{x}}_{\beta}\, . 
\end{equation}
Let us now introduce the matrix $\mathbf{c}(t)$ given by~\eqref{eq:cab} to rewrite
\begin{equation}
\label{eq:GLHGlinalpha2bApp}
\frac{d\delta\hat{\mathbf{x}}_{\beta}}{dt}+ \sum_\alpha c_{\alpha\beta}\delta\hat{\mathbf{x}}_{\alpha}= \mathbf{J}_\mathbf{F}(\mathbf{s})\delta\hat{\mathbf{x}}_{\beta}+\varepsilon \Lambda^{(\beta)}\mathbf{J}_\mathbf{H}(\mathbf{s})\delta\hat{\mathbf{x}}_{\beta}\, ,
\end{equation}
and recalling that $c_{\alpha\beta}=-c_{\beta\alpha}$ we eventually get:
\begin{equation}
\label{eq:GLHGlinalpha3App}
\frac{d\delta\hat{\mathbf{x}}_{\beta}}{dt}(t) = \sum_\alpha c_{\beta\alpha}(t)\delta\hat{\mathbf{x}}_{\alpha}(t)+\left[\mathbf{J}_\mathbf{F}(\mathbf{s}(t))+\varepsilon \Lambda^{(\beta)}(t)\mathbf{J}_\mathbf{H}(\mathbf{s}(t))\right]\delta\hat{\mathbf{x}}_{\beta}(t)\, . 
\end{equation}

The whole system state is described by $\delta\hat{\mathbf{x}}=(\delta\hat{\mathbf{x}}^\top_{1},\dots,\delta\hat{\mathbf{x}}_{n}^\top)^\top$. By introducing the Kronecker product, $\otimes$, one can rewrite the previous equation as
\begin{equation}
\label{eq:GLHGlinalpha3compactApp}
\frac{d\delta\hat{\mathbf{x}}}{dt} =  \left[\mathbf{c}\otimes \mathbf{1}_d+\mathbf{1}_n\otimes \mathbf{J}_\mathbf{F}+\varepsilon \mathbf{\Lambda}\otimes\mathbf{J}_\mathbf{H}\right]\delta\hat{\mathbf{x}}:=\mathbf{M}\delta\hat{\mathbf{x}}\, ,
\end{equation}
where we denoted by $\mathbf{\Lambda}$ the diagonal matrix containing the eigenvalues $\Lambda^{(\alpha)}$ on the diagonal.  Let us observe that because of the Perron effect, the negativity (resp. positivity) of the eigenvalues of the matrix $\mathbf{M}$ defined by the right hand side of the latter equation, does not allow to  conclude on the stability (resp. instability) of the reference solution~\cite{Perron1930,Vinograd1952}.

\section{Turing instability on time varying network. Application to the Brusselator.}
\label{sec:ti}

The aim of this section is to show how the formalism of Turing instability can be included in the description given by Eq.~\eqref{eq:maineq} and the analysis resulting from Eq.~\eqref{eq:GLHGlinalpha3compact}. The first step is to assume a linear coupling $\mathbf{H}(\mathbf{x})=\mathbf{D}\mathbf{x}$. Moreover in absence of cross-diffusion the latter takes a diagonal form with positive entries $\mathbf{D}=\mathrm{diag}(D_1,\dots,D_d)$. Finally, the Turing instability is often studied in a $2$ dimensional setting, i.e., $d=2$. In conclusion Eq.~\eqref{eq:maineq} can be rewritten as
 \begin{equation*}
\frac{d\mathbf{x}_i}{dt}=\mathbf{F}(\mathbf{x}_i) + \sum_{j} L_{ij}(t) \mathbf{D} \mathbf{x}_j\, ,
\end{equation*}
where we recall that the system state of the $i$-th unit is described by $\mathbf{x}_i=(u_i,v_i)$. Let us notice that the parameter $\varepsilon$ has been incorporated into the matrix $\mathbf{D}$. 

The second key ingredient is the assumption that the reference solution has to be stationary, $\mathbf{s}(t)=\mathbf{s}_0$, and stable with respect to homogeneous perturbation. The Turing instability is obtained once the reference solution becomes unstable for a suitable choice of the matrix $\mathbf{D}$, of the model parameters and of the topology of the underlying network.

By setting $\mathbf{F}(\mathbf{x}_i)=(f(u_i,v_i),g(u_i,v_i))$ and $\mathbf{D}=\mathrm{diag}(D_u,D_v)$ the latter equation returns Eq.~\eqref{eq:Tnet}. The conditions for the stability of the homogeneous equilibrium $\mathbf{s}_0=(u^*,v^*)$ are $\mathrm{tr}(\mathbf{J}_0)<0$ and $\det(\mathbf{J}_0)>0$, where $\mathbf{J}_0$ is the Jacobian of the reaction part evaluated on such equilibrium.

The condition for the  onset of a Turing instability can be checked by performing a linear stability analysis of the complete system. We thus set $\delta x_i=u_i-u^*$, $\delta y_i=v_i-v^*$ and  linearise~\eqref{eq:Tnet} about the equilibrium, to get
\begin{equation}
\label{eq:Tnetlin}
\begin{dcases}
\frac{d\delta x_i}{dt}&=\partial_uf \delta x_i+\partial_vf \delta y_i+D_u\sum_{j=1}^{n}L_{ij}(t) \delta x_j  \\ 
\frac{d\delta y_i}{dt}&=\partial_ug \delta x_i+\partial_vg \delta y_i+D_v\sum_{j=1}^{n}L_{ij}(t) \delta y_j 
\end{dcases} \quad\forall i=1,\dots,n\, ,
\end{equation}
where the partial derivatives have been computed at the equilibrium. To proceed in the analysis we develop, as done above, the perturbation on the basis of eigenvectors of the Laplace matrix, that is
\begin{equation}
\label{eq:proj}
\delta x_i(t)=\sum_\alpha \delta\hat{x}_\alpha(t) \phi^{(\alpha)}_i(t)\text{ and }\delta y_i(t)=\sum_\alpha \delta\hat{y}_\alpha(t) \phi^{(\alpha)}_i(t)\quad \forall i=1,\dots,n\, .
\end{equation}
Inserting this ansatz into Eq.~\eqref{eq:Tnetlin} we get (to lighten the notation we did not explicitly write the time dependence of the variables)
\begin{equation}
\label{eq:Tnetlin2}
\begin{dcases}
\sum_\alpha \frac{d\delta\hat{x}_\alpha}{dt} \phi^{(\alpha)}_i+\sum_\alpha \delta\hat{x}_\alpha \frac{d \phi^{(\alpha)}_i}{dt}&=\partial_u f \sum_\alpha \delta\hat{x}_\alpha \phi^{(\alpha)}_i +\partial_v f \sum_\alpha \delta\hat{y}_\alpha \phi^{(\alpha)}_i+D_u\sum_{j=1}^{n}L_{ij} \sum_\alpha \delta\hat{x}_\alpha\phi^{(\alpha)}_j  \\ 
\sum_\alpha \frac{d\delta\hat{y}_\alpha}{dt} \phi^{(\alpha)}_i+\sum_\alpha \delta\hat{y}_\alpha\frac{d \phi^{(\alpha)}_i}{dt}&=\partial_ug \sum_\alpha \delta\hat{x}_\alpha \phi^{(\alpha)}_i+\partial_vg \sum_\alpha \delta\hat{y}_\alpha \phi^{(\alpha)}_i+D_v\sum_{j=1}^{n}L_{ij} \sum_\alpha \delta\hat{y}_\alpha\phi^{(\alpha)}_j
\end{dcases}\, ,
\end{equation}
for all $i=1,\dots,n$. By using the definition of $\sum_jL_{ij} \phi^{(\alpha)}_j=\Lambda^{(\alpha)}\phi^{(\alpha)}_i$, by projecting on the eigenvector $\vec{\phi}^{(\beta)}$ and by recalling the definition of the matrix $\mathbf{c}$ given by Eq.~\eqref{eq:cab}, we get
\begin{equation}
\label{eq:Tnetlin4}
\begin{dcases}
\frac{d\delta\hat{x}_\beta}{dt}&=\sum_\alpha c_{\beta\alpha}\delta\hat{x}_\alpha +\partial_u f \delta\hat{x}_\beta +\partial_v f  \delta\hat{y}_\beta+D_u  \Lambda^{(\beta)}  \delta\hat{x}_\beta  \\ 
\frac{d\delta\hat{y}_\beta}{dt} &=\sum_\alpha c_{\beta\alpha}\delta\hat{y}_\alpha +\partial_ug  \delta\hat{x}_\beta +\partial_vg  \delta\hat{y}_\beta +D_v \Lambda^{(\beta)}  \delta\hat{y}_\beta
\end{dcases} \quad\forall \beta=1,\dots,n\, .
\end{equation}

Let us observe that at variance from for the classical Turing setting, we cannot decouple the original linear system into $n$ systems, each referred  to a single mode. Indeed, matrix $\mathbf{c}$ mixes the modes and their contribution.

For a sake of definitiveness, let us consider the Brusselator model defined on the small time varying network previously defined. This amounts to set $f(u,v)=1-(b+1)u+cu^2v$ and $g(u,v)=bu-cu^2v$, where $b$ and $c$ are the positive model parameters. The stationary equilibrium is thus $u^* = 1$ and $v^* = b/c$, while the Jacobian of the reaction part evaluated on the equilibrium is $\partial_u f = b-1$, $\partial_vf = c$, $\partial_u g = -b$ and $\partial_v g = -c$. The condition for the stability of the homogeneous equilibrium is thus $\mathrm{tr}(\mathbf{J}_0)=b-1-c<0$ and $\mathrm{det}(\mathbf{J}_0)=c>0$.

Consider thus Eq.~\eqref{eq:Tnetlin4} for the possible values of $\beta$. The case $\beta=1$ is straightforward, being $c_{1\alpha}=0$ for all $\alpha$ and $\Lambda^{(1)}=0$, it returns the decoupled system
\begin{equation*}
\begin{dcases}
\frac{d\delta\hat{x}_1}{dt}&=\partial_u f \delta\hat{x}_1 +\partial_v f  \delta\hat{y}_1 \\ 
\frac{d\delta\hat{y}_1}{dt} &=\partial_ug \delta\hat{x}_1 +\partial_vg  \delta\hat{y}_1\, ,
\end{dcases}
\end{equation*}
The condition for the stability of the homogeneous equilibrium, $(u^*,v^*)$, i.e., $\partial_u f +\partial_v f<0$ and $\partial_u f \partial_v g-\partial_v f\partial_u g>0$, ensures that $\delta\hat{x}_1(t)\rightarrow 0$ and $\delta\hat{y}_1(t)\rightarrow 0$.

The remaining modes, $\beta=2$ and $\beta=3$, do satisfy (recall that $\Lambda^{(2)}=-1$ and $\Lambda^{(3)}=-2$):
\begin{equation*}
\begin{dcases}
\frac{d\delta\hat{x}_2}{dt}&=\Omega \delta\hat{x}_3 +\partial_u f \delta\hat{x}_2 +\partial_v f  \delta\hat{y}_2-D_u \delta\hat{x}_2  \\ 
\frac{d\delta\hat{y}_2}{dt} &=\Omega \delta\hat{y}_3 +\partial_ug  \delta\hat{x}_2 +\partial_vg  \delta\hat{y}_2 -D_v \delta\hat{y}_2\\
\frac{d\delta\hat{x}_3}{dt}&=-\Omega \delta\hat{x}_2 +\partial_u f \delta\hat{x}_3 +\partial_v f  \delta\hat{y}_3-2D_u \delta\hat{x}_3 \\ 
\frac{d\delta\hat{y}_3}{dt} &=-\Omega \delta\hat{y}_2 +\partial_ug  \delta\hat{x}_3 +\partial_vg  \delta\hat{y}_3 -2D_v \delta\hat{y}_3\, .
\end{dcases}
\end{equation*}
To determine the (in)stability character of the solution we have to compute the eigenvalues of the linear system and consider the one with the largest real part. If this latter results to be positive, then the homogenous equilibrium is destabilised otherwise it remains stable and the perturbation fades away.

\section{Synchronisation on time varying network. Application to the Stuart-Landau model}
\label{sec:synchSL}
Let us consider a system made of $n$ identical Stuart-Landau (SL) oscillators. It can be cast in the framework of Eq.~\eqref{eq:maineq} by observing that the SL is a $d=1$ dimensional but complex system. Thus by setting $\mathbf{x}_j=w_j$, i.e., the complex amplitude, $\mathbf{F}(w)=\sigma w-\beta w|w|^2$ and $\mathbf{H}(w)=w|w|^{m-1}$, we eventually obtain Eq.~\eqref{eq:maineqSL} in the main text.

We assume the parameters to ensure that the isolated SL converges (that is $\sigma_\Re>0$ and $\beta_\Re>0$) to the limit cycle solution, $\hat{z}(t)=\sqrt{\sigma_\Re/\beta_\Re}e^{i\omega t}$, where $\omega=\sigma_\Im-\beta_\Im \sigma_\Re/\beta_\Re$. To study its stability with respect to heterogeneous perturbation we introduce two functions, $\rho_j(t)$ and $\theta_j(t)$, and we rewrite the complex amplitude as follows
\begin{equation}
\label{eq:wpertApp}
w_j(t)=\hat{z}(t)(1+\rho_j(t))e^{i\theta_j(t)}\, .
\end{equation}
Let us observe that invoking the smallness of $\rho_j(t)$ and $\theta_j(t)$ the previous equation can be rewritten as
\begin{equation}
\label{eq:wpert2App}
w_j(t)=\hat{z}(t)(1+\rho_j(t)+i\theta_j(t))+\mathrm{h.o.t.}\Rightarrow w_j(t)-\hat{z}(t) = \hat{z}(t)(\rho_j(t)+i\theta_j(t))+\mathrm{h.o.t.}=\delta w_j \, ,
\end{equation}
that measures  the distance with respect to the reference solution as done in the main text. In the same limit, the nonlinear coupling reduces to
\begin{equation*}
 H(w_\ell)=\hat{z}\left(\frac{\sigma_\Re}{\beta_{\Re}}\right)^{\frac{m-1}{2}}\left(1+m\rho_\ell+i\theta_\ell\right)+\mathrm{h.o.t.}\, .
\end{equation*}

Inserting the relation~\eqref{eq:wpertApp} into Eq.~\eqref{eq:maineqSL}, by performing a Taylor expansion to first order and by separating the real and the imaginary parts of $w_j$, one can obtain two ODEs ruling the time evolution of $\rho_j$ and $\theta_j$
\begin{equation}
\label{eq:maineqSLlin}
\begin{dcases}
\frac{d\rho_j}{dt} = -2\sigma_\Re \rho_j  +\left(\frac{\sigma_\Re}{\beta_{\Re}}\right)^{\frac{m-1}{2}}\sum_\ell L_{j\ell}(t)\left(m\mu_\Re \rho_\ell-\mu_\Im \theta_\ell\right)\\
\frac{d\theta_j}{dt} = -2\beta_\Im\frac{\sigma_\Re}{\beta_{\Re}} \rho_j  +\left(\frac{\sigma_\Re}{\beta_{\Re}}\right)^{\frac{m-1}{2}}\sum_\ell L_{j\ell}(t)\left(m\mu_\Im \rho_\ell+\mu_\Re \theta_\ell\right)\, .
\end{dcases}
\end{equation}

This is a linear non-autonomous system. The information on the network evolution is stored in the Laplace matrix $L_{ij}(t)$. We can again decompose $\rho_j(t)$ and $\theta_j(t)$ on the eigenbasis $\vec{\phi}^{(\alpha)}(t)$, $\alpha=1,\dots,n$
\begin{equation*}
 \rho_j=\sum_\alpha \hat{\rho}_\alpha \phi^{(\alpha)}_j \text{ and } \theta_j=\sum_\alpha \hat{\theta}_\alpha \phi^{(\alpha)}_j \, ,
\end{equation*}
to eventually obtain the equation analogous of Eq.~\eqref{eq:GLHGlinalpha3App}
\begin{equation}
\label{eq:maineqSLlinMSFApp}
\begin{dcases}
\frac{d{\hat{\rho}}_\gamma}{dt} = \sum_\alpha c_{\gamma\alpha}(t)\hat{\rho}_\alpha-2\sigma_\Re \hat{\rho}_\gamma  +\left(\frac{\sigma_\Re}{\beta_{\Re}}\right)^{\frac{m-1}{2}}\Lambda^{(\gamma)}(t)\left(m\mu_\Re \hat{\rho}_\gamma-\mu_\Im \hat{\theta}_\gamma\right)\\
\frac{d{\hat{\theta}}_\gamma}{dt} = \sum_\alpha c_{\gamma\alpha}(t)\hat{\theta}_\alpha-2\beta_\Im\frac{\sigma_\Re}{\beta_{\Re}} \hat{\rho}_\gamma  +\left(\frac{\sigma_\Re}{\beta_{\Re}}\right)^{\frac{m-1}{2}}\Lambda^{(\gamma)}(t)\left(m\mu_\Im \hat{\rho}_\gamma+\mu_\Re \hat{\theta}_\gamma\right)\, .
\end{dcases}
\end{equation}

For the sake of definitiveness let us consider the coupling being defined by the small network presented above. The general system~\eqref{eq:maineqSLlinMSFApp} returns thus for mode $\gamma=1$ 
\begin{equation*}
 \begin{dcases}
\frac{d{\hat{\rho}}_1}{dt} = -2\sigma_\Re\hat{\rho}_1\\
\frac{d{\hat{\theta}}_1}{dt} = -2\beta_\Im\frac{\sigma_\Re}{\beta_{\Re}} \hat{\rho}_1\, ,
\end{dcases}
\end{equation*}
whose solutions are $\hat{\rho}_1(t)= e^{-2\sigma_\Re t}\hat{\rho}_1(0)$ and $\hat{\theta}_1(t)=\hat{\theta}_1(0) + \frac{\beta_\Im}{\beta_{\Re}}\hat{\rho}_1(0)\left(e^{-2\sigma_\Re t}-1 \right)$ and thus $\hat{\rho}_1(t)\rightarrow 0$ (recall $\sigma_\Re>0$) and $\hat{\theta}_1(t)\rightarrow \theta_1(0)-\beta_\Im \hat{\rho}_1(0)/\beta_\Re$.

The remaining modes $\gamma=2$ and $\gamma=3$ return (recall that $\Lambda^{(2)}=-1$ and $\Lambda^{(3)}=-2$)
\begin{equation*}
\begin{dcases}
\frac{{\hat{\rho}}_2}{dt} = \Omega\hat{\rho}_3-2\sigma_\Re \hat{\rho}_2  -\left(\frac{\sigma_\Re}{\beta_{\Re}}\right)^{\frac{m-1}{2}}\left(m\mu_\Re \hat{\rho}_2-\mu_\Im \hat{\theta}_2\right)\\
\frac{d{\hat{\rho}}_3}{dt} = -\Omega\hat{\rho}_2-2\sigma_\Re \hat{\rho}_3  -2\left(\frac{\sigma_\Re}{\beta_{\Re}}\right)^{\frac{m-1}{2}}\left(m\mu_\Re \hat{\rho}_3-\mu_\Im \hat{\theta}_3\right)\\
\frac{d{\hat{\theta}}_2}{dt} = \Omega\hat{\theta}_3-2\beta_\Im\frac{\sigma_\Re}{\beta_{\Re}} \hat{\rho}_2  -\left(\frac{\sigma_\Re}{\beta_{\Re}}\right)^{\frac{m-1}{2}}\left(m\mu_\Im \hat{\rho}_2+\mu_\Re \hat{\theta}_2\right)\\
\frac{d{\hat{\theta}}_3}{dt} = -\Omega\hat{\theta}_2-2\beta_\Im\frac{\sigma_\Re}{\beta_{\Re}} \hat{\rho}_3  -2\left(\frac{\sigma_\Re}{\beta_{\Re}}\right)^{\frac{m-1}{2}}\left(m\mu_\Im \hat{\rho}_3+\mu_\Re \hat{\theta}_3\right)\, .
\end{dcases}
\end{equation*}

The synchronisation is thus obtained if all the eigenvalues of the latter system have negative real part, while if at least one eigenvalue exists with positive real part, then the system desynchronises and converges to a (generically) heterogeneous solution.

\begin{figure*}[ht]
\centering
\includegraphics[scale=0.24]{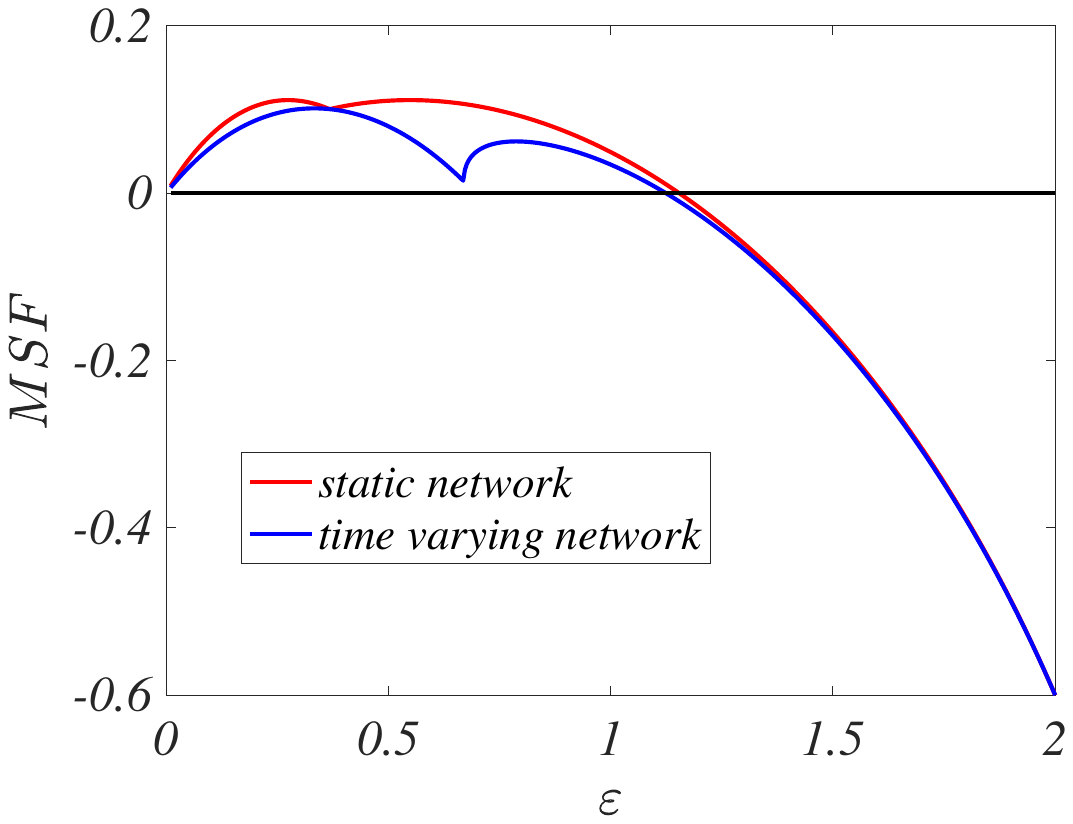}\quad\includegraphics[scale=0.24]{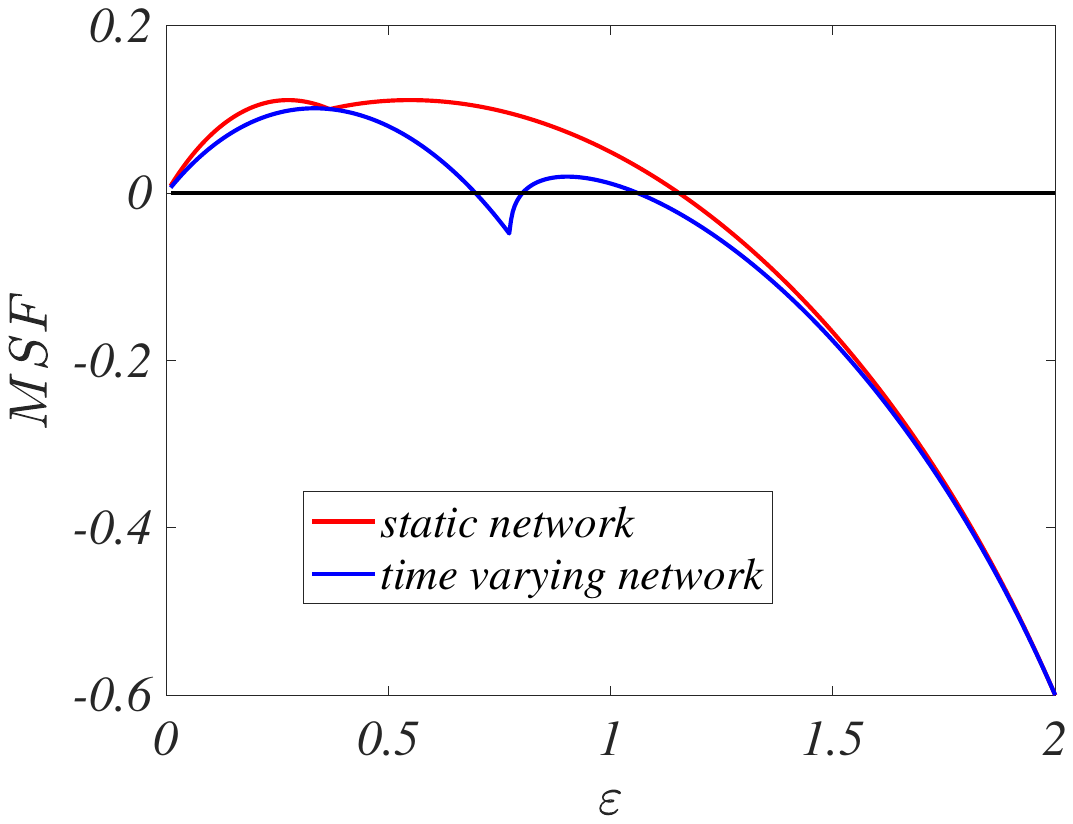}
\vspace{-1.5cm}
\caption{\textbf{Peculiar behaviour of the MSF for the Stuart-Landau nonlinearly coupled oscillators.} We report the MSF for two values of $\Omega$, one below the abrupt transition (left panel) and one above the same transition (right panel). The remaining parameters have been fixed to the values, $\sigma = 1.0+4.3i$, $\beta= 1.0+1.1i$, $\mu_0 = 0.1-0.5i$ and $m = 3$.}
\label{fig:fig3SL}
\end{figure*}

In Fig.~\ref{fig:fig2SL} we have shown the position of the nontrivial zeros of the MSF, $\varepsilon_0(\Omega)$, as a function of $\Omega$, and we have observed the existence of an interval of values $I=[\Omega_1,\Omega_2]$, $\Omega_1\sim 0.106$ and $\Omega_2\sim 0.162$, such that for all $\Omega\in I$ there exist three values of $\varepsilon_0(\Omega)$. The particular shape of MSF for the SL model can explain this behaviour. Indeed as $\Omega\rightarrow 0$ the blue curve (see Fig.~\ref{fig:fig1SL}) tends toward the red one, however for large $\Omega$ the blue curve presents a ``cusp'' and thus three zeros can emerge if the cusp is ``deep enough'' (see Fig.~\ref{fig:fig3SL}). The existence of such interval is interesting because we can find a ``window'' in $\Omega$ for which the time varying network synchronises, while the static one does not.

\begin{figure*}[ht]
\centering
\includegraphics[scale=0.24]{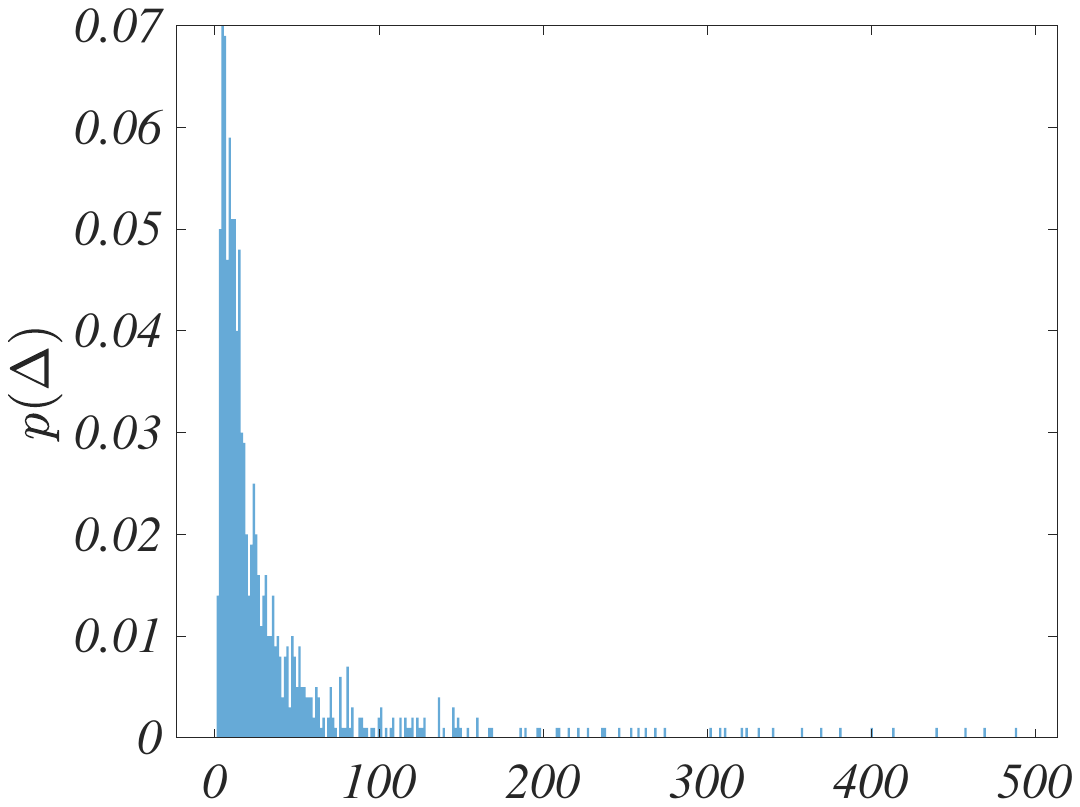}\quad\includegraphics[scale=0.24]{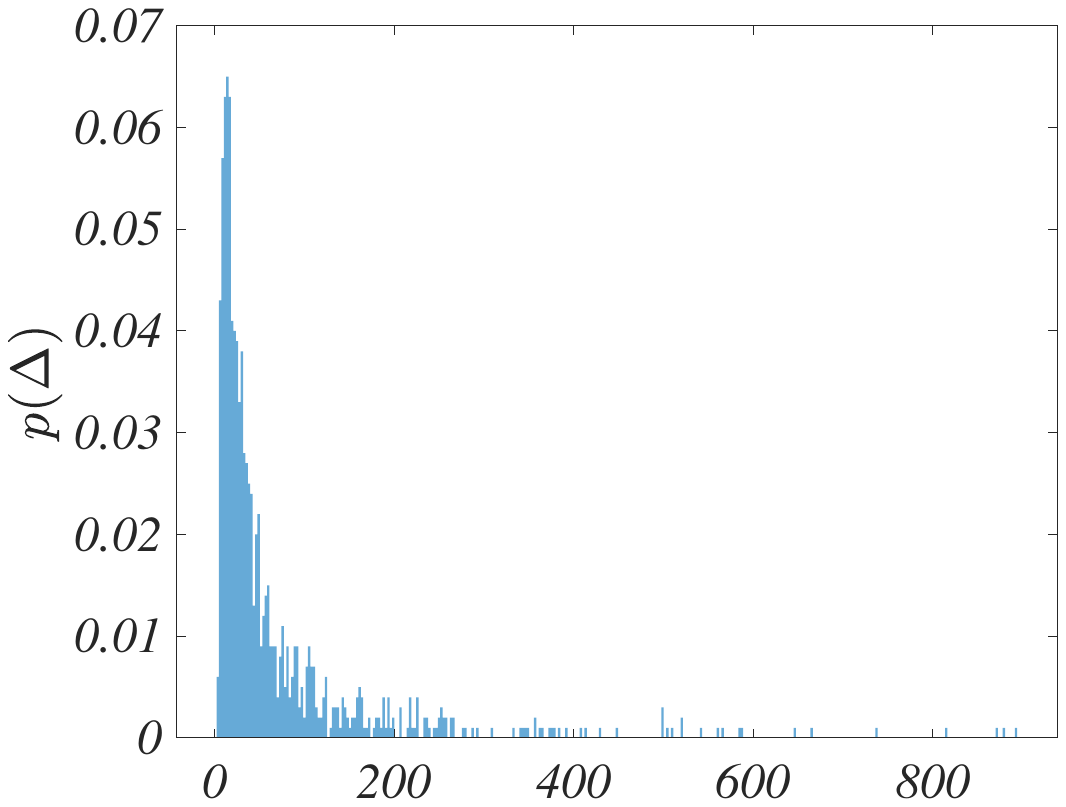}
\vspace{-1.5cm}
\caption{\textbf{Synchronisation threshold.} We report the pdf of $\Delta(\mathbf{c}):=\varepsilon_0(\mathbf{0})-\varepsilon_0(\mathbf{c})$ for $1000$ networks composed by $50$ nodes (left panel) and $100$ nodes (right panel).}
\label{fig:DeltacSL}
\end{figure*}

In the main text we have shown that the simple $3$-nodes network allows an interval of coupling strength for which the SL synchronises that is larger than once assume a static network. Let us conclude this section by showing that this results holds true for networks of arbitrary size built using a constant skew symmetric matrix $\mathbf{c}$, as done in the case of the small network with three nodes. To support this claim we build $1000$ skew symmetric matrices $\mathbf{c}$, whose first row and column are zeros. For each matrix we computed the MSF for the SL defined on top of the networks obtained using such matrix and by assigning also a random nonpositive set of eigenvalues; eventually we determined the smallest non trivial zero of the MSF, $\varepsilon_0(\mathbf{c})$. We computed also the same quantity in the case of a null matrix, corresponding thus to a static network, $\varepsilon_0(\mathbf{0})$, with the same random eigenvalues. Let $\Delta(\mathbf{c}):=\varepsilon_0(\mathbf{0})-\varepsilon_0(\mathbf{c})$, then we observed that $\Delta(\mathbf{c})>0$, implying that the SL defined on top of time varying network always exhibits a larger range of coupling strength associated to synchronisation. In Fig.~\ref{fig:DeltacSL} we report the probability distribution function (pdf) of $\Delta(\mathbf{c})$ for networks of $50$ nodes (left panel) and $100$ nodes (right panel). One can observe that both distributions are broad, which implies the existence of static networks requiring an extremely large coupling strength to synchronise. Moreover the minimum value of $\Delta(\mathbf{c})$ is $\sim 1.17$ once considering networks of $50$ nodes, while it raises to $\sim 2.31$ for $100$ nodes, this implies that SL defined on top of large stationary networks are more difficult to synchronise than in the case of time varying networks.

\section{Synchronisation of chaotic trajectories on time varying network. The Lorenz model.}
\label{sec:synchLorenz}
As previously stated, the results hereby presented are of general nature and go beyond the specific examples we exposed. To support this claim, let us conclude this section by presenting an application of the above theory to the phenomenon of synchronisation of chaotic trajectories on time varying networks and to stress the central role played by the matrix $\mathbf{c}(t)$ in the generalised MSF~\eqref{eq:GLHGlinalpha3compact}.

Without loss of generality we will use  for a demonstrative application the Lorenz model~\cite{Lorenz1963}
\begin{equation}
\label{eq:lorenz}
\begin{dcases}
 \dot{x}&=\sigma (y-x)\\
 \dot{y}&=x(\rho-z)-y\\
 \dot{z}&=xy-\beta z\quad .
\end{dcases}
\end{equation}
In the following we will fix the model parameters to the ``standard values'', $\beta=2$, $\sigma=10$ and $\rho=28$ for which the system exhibits the chaotic orbit with a ``butterfly shape''. We then consider $n$ identical copies of the above system, interacting through the simple $3$-nodes time varying network presented above, using a linear diffusive $1\rightarrow 1$ coupling, as defined in~\cite{HCLP}
\begin{equation}
\label{eq:lorenz2}
\begin{dcases}
 \dot{x}_i&=\sigma (y_i-x_i)+\varepsilon \sum_j L_{ij}(t)x_j\\
 \dot{y}_i&=x_i(\rho-z_i)-y_i\\
 \dot{z}_i&=x_iy_i-\beta z_i\quad .
\end{dcases}
\end{equation}

Let $\mathbf{s}(t)=(x_L(t),y_L(t),z_L(t))$ be the reference chaotic trajectory, we can define the deviation $\delta\hat{\mathbf{x}}=(\delta\hat{\mathbf{x}}^\top_{1},\dots,\delta\hat{\mathbf{x}}_{n}^\top)^\top$, where $\delta\hat{\mathbf{x}}^\top_{I}=\left(x_i(t)-x_L(t),y_i(t)-y_L(t),z_i(t)-z_L(t)\right)$, and eventually obtain the analogous of the MSF~\eqref{eq:GLHGlinalpha3}:
  \begin{equation*}
\frac{d\delta\hat{\mathbf{x}}}{dt} =  \left[\mathbf{c}\otimes \mathbf{1}_3+\mathbf{1}_n\otimes \mathbf{J}_\mathbf{F}+\varepsilon \mathbf{\Lambda}\otimes\mathbf{J}_\mathbf{H}\right]\delta\hat{\mathbf{x}}:=\mathbf{M}\delta\hat{\mathbf{x}}\, ,
\end{equation*}
where
\begin{equation*}
\mathbf{\Lambda}=\left(
\begin{matrix}
 0 & 0 & 0\\ 0& -1& 0\\ 0& 0 &-2
\end{matrix}\right)\, , \,
\mathbf{J}_\mathbf{F}(\mathbf{s}(t))=\left(
\begin{matrix}
 -\sigma & \sigma & 0\\ \rho-z_L(t)& -1& -x_L(t)\\ y_L(t)& x_L(t) &-\beta
\end{matrix}\right)\text{ and }\mathbf{J}_\mathbf{H}(\mathbf{s}(t))=\left(
\begin{matrix}
 1 & 0 & 0\\ 0& 0& 0\\ 0& 0 &0
\end{matrix}\right)\, .
\end{equation*}

\begin{figure*}[ht]
\centering
\includegraphics[scale=0.22]{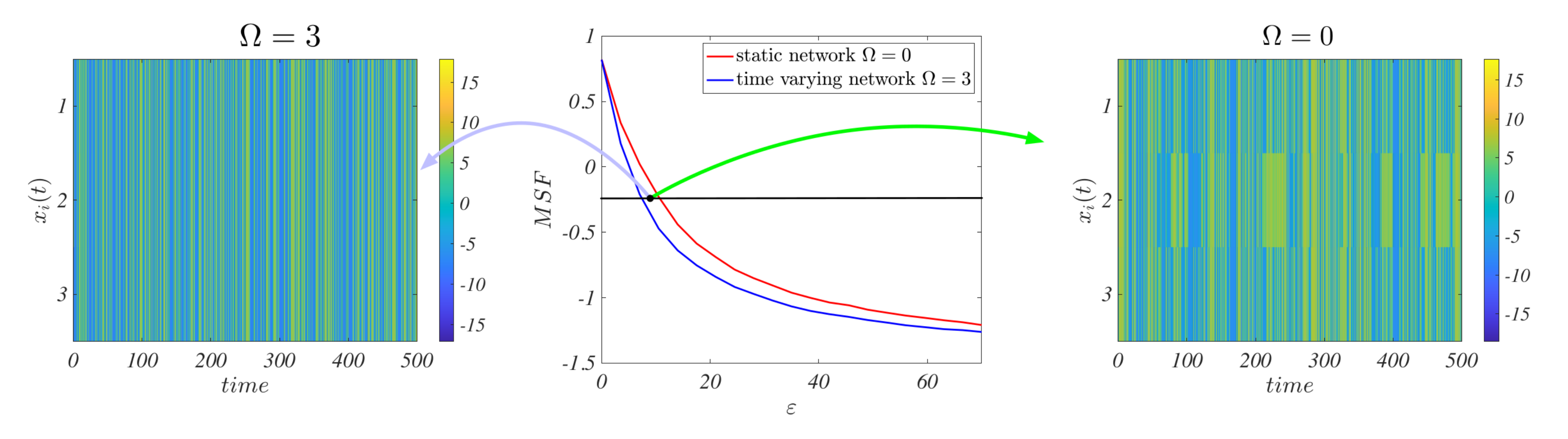}
%\vspace{-1.5cm}
\caption{\textbf{Synchronisation of Lorenz oscillators on time varying networks.} Middle panels report the MSF for the time varying network, $\Omega=3$, (blue curve) and the static one (red curve) as a function of the coupling strength $\varepsilon$. We can observe that the former has a zero for $\varepsilon \sim 5.12$ while the latter for $\varepsilon \sim 7.31$; this implies the existence of an interval of coupling strength for which the Lorenz trajectories synchronise once coupled via the time varying network (left panel), while they do not on the static network (right panel). The displayed pictures ($x_i$ vs. time) are obtained for $\varepsilon=6.0$.}
\label{fig:resultsLorenz}
\end{figure*}

Solving numerically the previous non-autonomous linear system one can compute the associated Master Stability Function (as introduced in the main body of the paper) as a function of the coupling strength $\varepsilon$, for both a static and time varying network. To this end we will consider the three nodes network introduced above. Results reported in Fig.~\ref{fig:resultsLorenz} show that the synchronisation depends on the matrix $\mathbf{c}$. Indeed, the parameter $\varepsilon$ can be chosen in such a way that synchrony is achieved on a time varying network while trajectories are found to behave differently (and hence patterns develop) on its static analogue. As already shown for the Stuart-Landau, the small $3$-nodes time varying network seems to enhance the synchronisation.

\begin{figure*}[ht]
\centering
\includegraphics[scale=0.33]{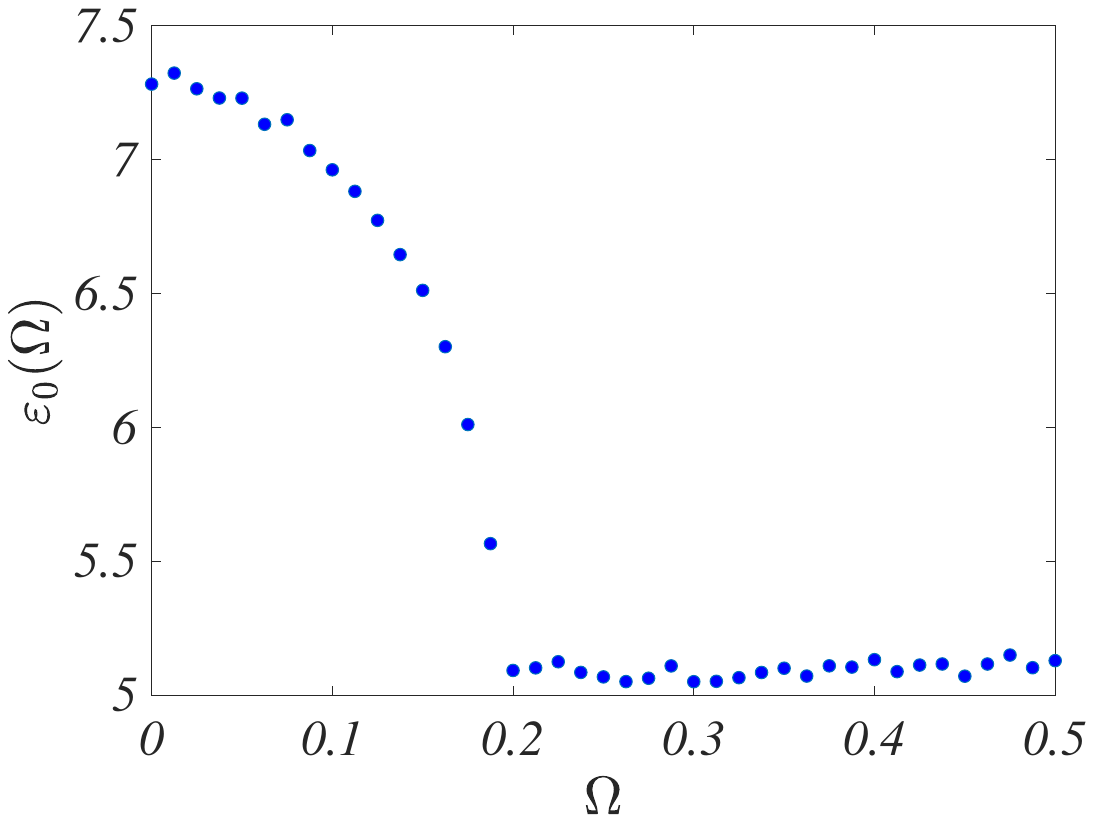}
\vspace{-2cm}
\caption{\textbf{Synchronisation of Lorenz oscillators on time varying networks.} On the left panel reports the MSF for the time varying network, $\Omega=0.2$, (blue curve) and the static one (red curve) as a function of the coupling strength $\varepsilon$. We can observe that the former has a first zero for $\varepsilon \sim 5.2$ while the latter for $\varepsilon \sim 7.3$; this implies the existence of an interval of coupling strength for which the Lorenz trajectories synchronise once coupled via the time varying network, while they do not on the static network. On the right panel we report the zero of the MSF, $\varepsilon_0(\Omega)$ as a function of $\Omega$.}
\label{fig:resultsLorenz2}
\end{figure*}

To conclude let us consider again the synchronisation of the chaotic solution of the Lorenz model as a function of the parameter $\varepsilon$, i.e., the strength of the coupling. Results reported in Fig.~\ref{fig:resultsLorenz2} show once again that $\varepsilon_0(0)>\varepsilon_0(\Omega)$ for the considered $\Omega$ allowing to conclude that the Lorenz system defined on a time varying network can synchronise more easily than the static network. Observe that now the MSF is convex, namely its second derivative with respect to the coupling strength is positive, and still the dynamic network ``more easily'' synchronises than the static one; we can thus conclude that the assumption about the convexity of the MSF used in~\cite{ZS2021} is not in general required.

Let us also note that the data are a bit noisy with respect to the SL case studied in the main text, the reason being that for the Lorenz case the MSF should be computed numerically by solving the nonlinear system and its linearised version for the deviation vector, while in the case of SL the MSF can be computed analytically as the involved matrix is constant.

\end{document}